\documentclass[12pt,preprint]{aastex}
\usepackage{graphics}
\usepackage{epsfig}

\newcommand{\ltsimeq}{\la}

\newcommand{\msun}{M$_{\odot}$}

\received{}
\revised{}
\accepted{}
\shortauthors{McQuinn et al.}
\shorttitle{The Duration of Starbursts}

\begin{document}
\title{The Nature of Starbursts : II. The Duration of Starbursts in Dwarf Galaxies\footnote{Based on observations made with the NASA/ESA Hubble Space Telescope, obtained from the Data Archive at the Space Telescope Science Institute, which is operated by the Association of Universities for Research in Astronomy, Inc., under NASA contract NAS 5-26555.}}
\author{Kristen B.~W. McQuinn\altaffilmark{1}, 
Evan D. Skillman\altaffilmark{1},
John M. Cannon\altaffilmark{2},
Julianne Dalcanton\altaffilmark{3},
Andrew Dolphin\altaffilmark{4},
Sebastian Hidalgo-Rodr\'{i}guez\altaffilmark{5},
Jon Holtzman\altaffilmark{6},
David Stark\altaffilmark{1},
Daniel Weisz\altaffilmark{1},
Benjamin Williams\altaffilmark{3}
}

\altaffiltext{1}{Department of Astronomy, School of Physics and
Astronomy, 116 Church Street, S.E., University of Minnesota,
Minneapolis, MN 55455, \ {\it kmcquinn@astro.umn.edu}} 
\altaffiltext{2}{Department of Physics and Astronomy, 
Macalester College, 1600 Grand Avenue, Saint Paul, MN 55105}
\altaffiltext{3}{Department of Astronomy, Box 351580, University 
of Washington, Seattle, WA 98195}
\altaffiltext{4}{Raytheon Company, 1151 E. Hermans Road, Tucson, AZ 85706}
\altaffiltext{5}{Instituto de Astrof\'{i}sica de Canarias, V\'{i}a L\'{a}ctea s/n.~E38200, La Laguna, Tenerife, Canary Islands, Spain}
\altaffiltext{6}{Department of Astronomy, New Mexico State University, Box 30001-Department 4500, 1320 Frenger Street, Las Cruces, NM 88003}

\begin{abstract}

The starburst phenomenon can shape the evolution of the host galaxy and the surrounding intergalactic medium. The extent of the evolutionary impact is partly determined by the duration of the starburst, which has a direct correlation with both the amount of stellar feedback and the development of galactic winds, particularly for smaller mass dwarf systems. We measure the duration of starbursts in twenty nearby, ongoing, and ``fossil" starbursts in dwarf galaxies based on the recent star formation histories derived from resolved stellar population data obtained with the \textit{Hubble Space Telescope}. Contrary to the shorter times of $3-10$ Myr often cited, the starburst durations we measure range from $450 - 650$ Myr in fifteen of the dwarf galaxies and up to 1.3 Gyr in four galaxies; these longer durations are comparable to or longer than the dynamical timescales for each system. The same feedback from massive stars that may quench the flickering SF does not disrupt the overall burst event in our sample of galaxies. While five galaxies present fossil bursts, fifteen galaxies show ongoing bursts and thus the final durations may be longer than we report here for these systems. One galaxy shows a burst that has been ongoing for only 20 Myr; we are likely seeing the beginning of a burst event in this system.  Using the duration of the starbursts, we calculate that the bursts deposited $10^{53.9}-10^{57.2}$ ergs of energy into the interstellar medium through stellar winds and supernovae and produced 3\%$-$26\% of the host galaxy's mass.

\end{abstract} 


\section{Characteristics of the Starburst Phenomenon \label{intro}}
Starbursts refer to periods of intense, massive star formation that persist on timescales much shorter than the Hubble time \citep{Searle1973}. The massive stars formed during a burst can alter the gas dynamics, future star formation, and chemical composition of the host galaxy through the stars' ionizing radiation, mass loss, and nucleosynthesis.  The evolutionary effects of starbursts can be dramatic and may be one of the keys to understanding the possible link between the evolution of dwarf irregular galaxies (dIrrs) to dwarf spheroidal galaxies (dSphs) \citep[e.g.,][]{Loose1986, Dekel1986, Silk1987, Davies1988} along with external effects such as ram pressure and tidal stripping from interactions with other systems \citep[e.g.,][]{Mayer2001a, Mayer2001b, vanZee2004, Mayer2006}. The strength of these feedback processes and their evolutionary impact depend on the SF activity and therefore are largely dictated by both the intensity and the duration of the starburst event. 

The duration of starbursts have been previously measured in individual galaxies \citep[e.g.,][]{Schaerer1999, Mas-Hesse1999, Tremonti2001} and in larger samples of galaxies \citep[e.g.,][]{Thornley2000, Harris2004}. The most commonly cited timescale for a starburst  is $5-10$ Myr, comparable to the lifetime of massive stars. It is logical to presume from the similarity of these timescales that bursts might be ``self-quenching''. In other words, the massive star formation of a burst produces stellar winds and supernovae explosions that could disrupt the gas fueling the star formation and extinguish the starburst on the timescale of a few to ten Myr. The hypothesis of ``self-quenching starbursts" has been put forth in a number of papers both observational \citep{Schaerer1999, Mas-Hesse1999, Thornley2000, Tremonti2001, Harris2004} and theoretical \citep{Tosi1989, Ferguson1998, Stinson2007}, all of which derive $\ltsimeq 10$ Myr durations for starbursts. This same short timescale has been adopted in a number of models. Ferguson et al.~(1998) use a burst duration of 10 Myr with a star formation efficiency of $10\%$ in modeling the redshift distribution of dwarf galaxies. Krueger et al.~(1995) assume a duration of 5 Myr when modeling the spectral energy distributions of bursting dwarf galaxies. The short duration assumed in models by \citet{Kauffmann1994} and by \citet{Hogg1997} affects the slope of the faint galaxy count in the galaxy luminosity function. 

However, our recent study of three nearby dwarf galaxies \citep{McQuinn2009} showed both that starbursts can last much longer (i.e., hundreds of Myr) and that starbursts are a global phenomenon that can migrate through a galaxy. The duration of starbursts is not set by the lifetime of massive stars in these galaxies. These results are in agreement with earlier suggestions that starbursts can last much longer than a few to ten Myr \citep{Calzetti1997, Greggio1998, Meurer2000}. A complimentary study by \citet{Schulte-Ladbeck2001} reported that although simulations support fluctuations in SFR on short timescales, long periods of quiescence between short bursts produce gaps in the simulated color$-$magnitude diagrams (CMDs) not seen in observational data. 

Starburst driven winds may feed the intergalactic medium (IGM) altering the internal kinematics and evolutionary path of the host galaxy. The importance of galactic winds and metal$-$rich gas outflows triggered by supernovae explosions in low mass galaxies has been much debated in the literature \citep[e.g.,][and references therein]{Dekel1986, Skillman1989, Skillman1997, Spaans1997, MacLow1999, Romano2006, Dalcanton2007, Martin2002, Tolstoy2009}. If starbursts are sustained on timescales of a few 100 Myr for the majority of dwarf starburst galaxies, longer bursts may more readily power galactic winds. 

Here, we directly measure the durations of starbursts in twenty nearby bursting dwarf galaxies from their star formation histories (SFHs) reconstructed from archival Hubble Space Telescope (HST) observations. We presented the details of the observations, data reduction, and method of reconstructing the SFHs in \citet[][hereafter, Paper I]{McQuinn2010}. In this paper, we provide a summary of the observations and data processing in \S\ref{obs}, measure the duration of the burst in each system in \S\ref{durations}, and compare the durations with the dynamical properties of each host galaxy in \S\ref{bursts}. The bursts' durations are compared  in \S\ref{flickering} with the $5-10$ Myr timescale often cited. Finally, we calculate the amount of stellar mass and energy produced in the burst in \S\ref{mass_energy}. Our conclusions are presented in \S\ref{conclusions}.

\section {The Sample Galaxies and Their Star Formation Histories \label{obs}}

\subsection {Observations and Photometry}
Our sample was selected from observations in the HST public archive based on three photometric conditions. First, we require both $V$ and $I$ band images of a galaxy in the archival observations. Second, we set a minimum photometric depth of the $I$ band observations of $\sim$2 mag below the tip of the red giant branch, which is required to accurately constrain the recent SFH of the galaxy \citep{Dolphin2002, Dohm-Palmer2002}. Third, the galaxies are required to lie close enough such that their stellar populations were resolved by HST imaging instruments. These properties were chosen to ensure robust reconstruction of the SFHs and measurements of the duration. The photometry for one galaxy, SBS~1415$+$437, stopped $\sim 0.5$ mag short of our depth requirement despite long integration times. We include the results on this well$-$studied galaxy in our sample for comparison purposes. The resulting sample of twenty galaxies is listed in Table~\ref{tab:durations}. 

All photometry used DOLPHOT or HSTphot \citep{Dolphin2000} on the standard HST pipeline processed and cleaned images. The details of the observations and photometric processing for eighteen of these galaxies are presented in Paper I and for two galaxies, DDO~165 and Ho~II, in \citet{Weisz2008}. As a representative example of the data quality and photometry, we present the imaging and corresponding CMD for one galaxy, UGC~9128, in the top two panels of Figure~\ref{fig:ugc9128}. The evolutionary features that drive the reconstruction of the SFHs can be readily identified in the CMD including the main sequence (MS) stars, blue and red helium burning (BHeB and RHeB respectively) stars, the red giant branch (RGB) stars, and the asymptotic giant branch (AGB) stars. The upper MS, the BHeB, and the RHeB stars are unambiguous signs of star formation in the past few 100 Myr. The helium burning (HeB) stars are of particular interest as a star in this stage of stellar evolution occupies a unique space in the CMD; the luminosity of a HeB star has a one$-$to$-$one correspondence to the mass of the star and, hence, the age of the star. Under the assumption of a universal initial mass function (IMF), this luminosity$-$mass$-$age relation of the HeB stars tightly constrains the SFH at recent times.

\subsection{SFH Method}
\begin{figure}
\epsscale{0.8}
\centering
\begin{tabular}{cc}
\epsfig{file=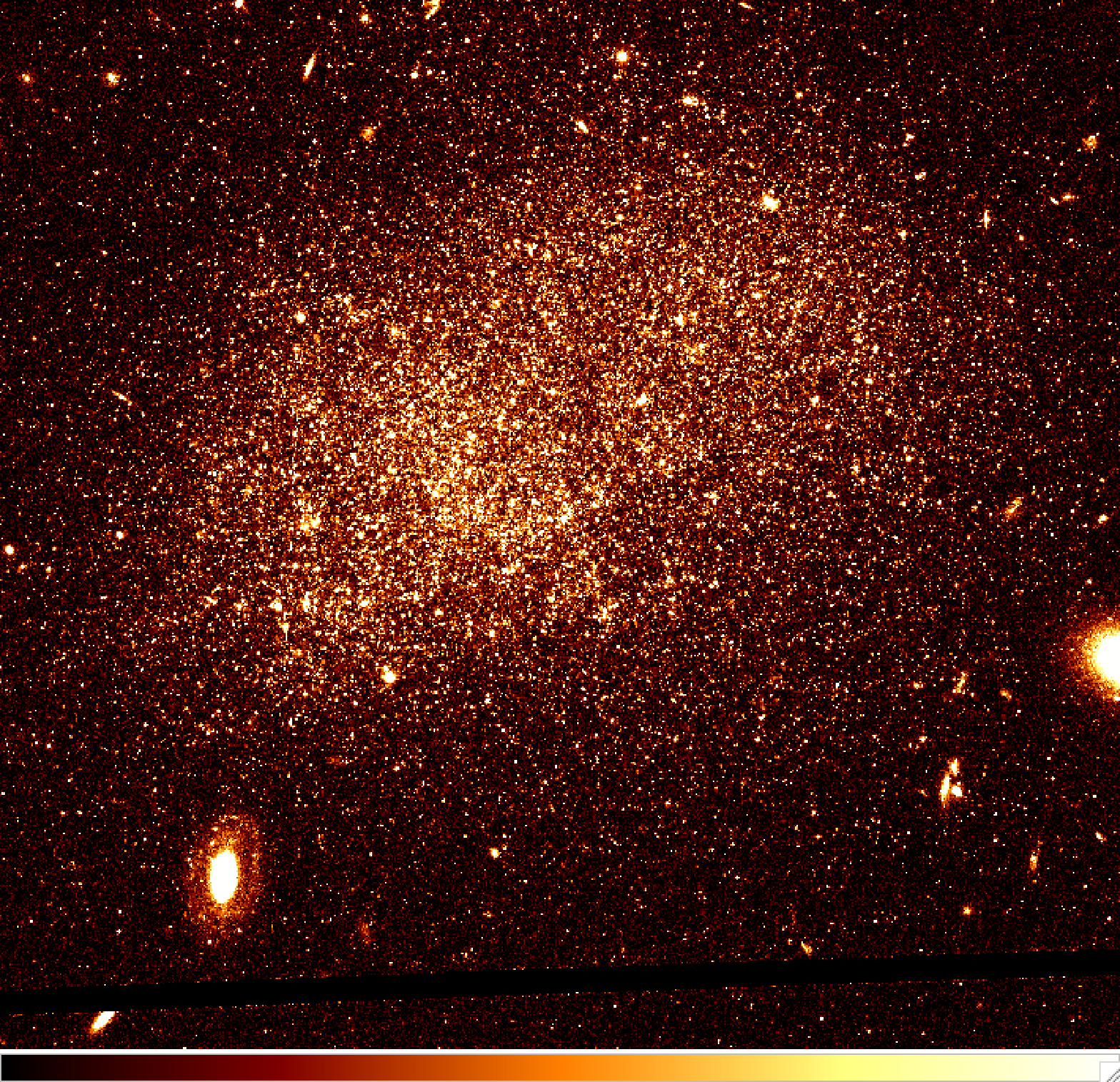,width=0.5\linewidth,clip=} & 
\epsfig{file=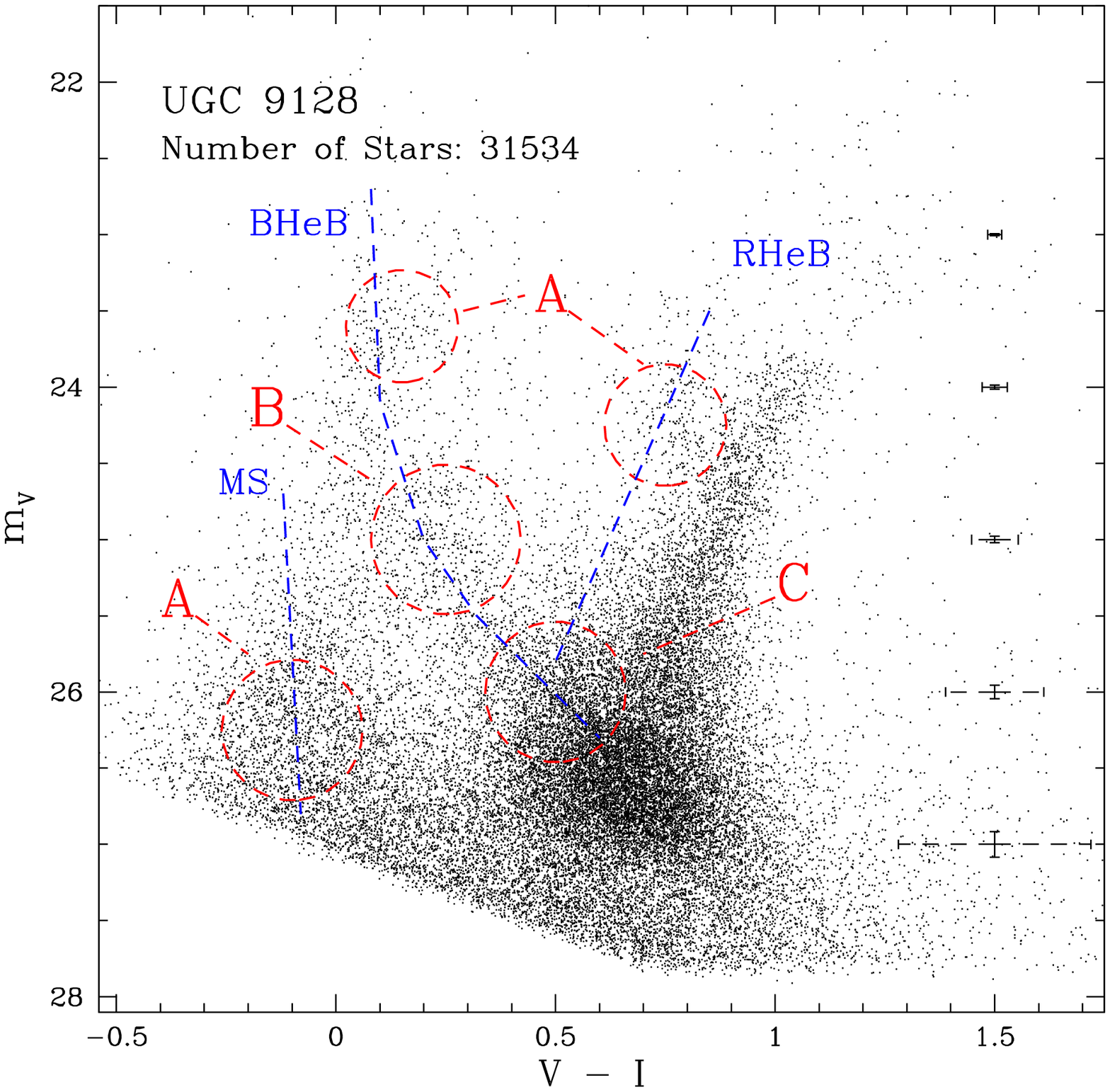,width=0.5\linewidth,clip=} \\
\epsfig{file=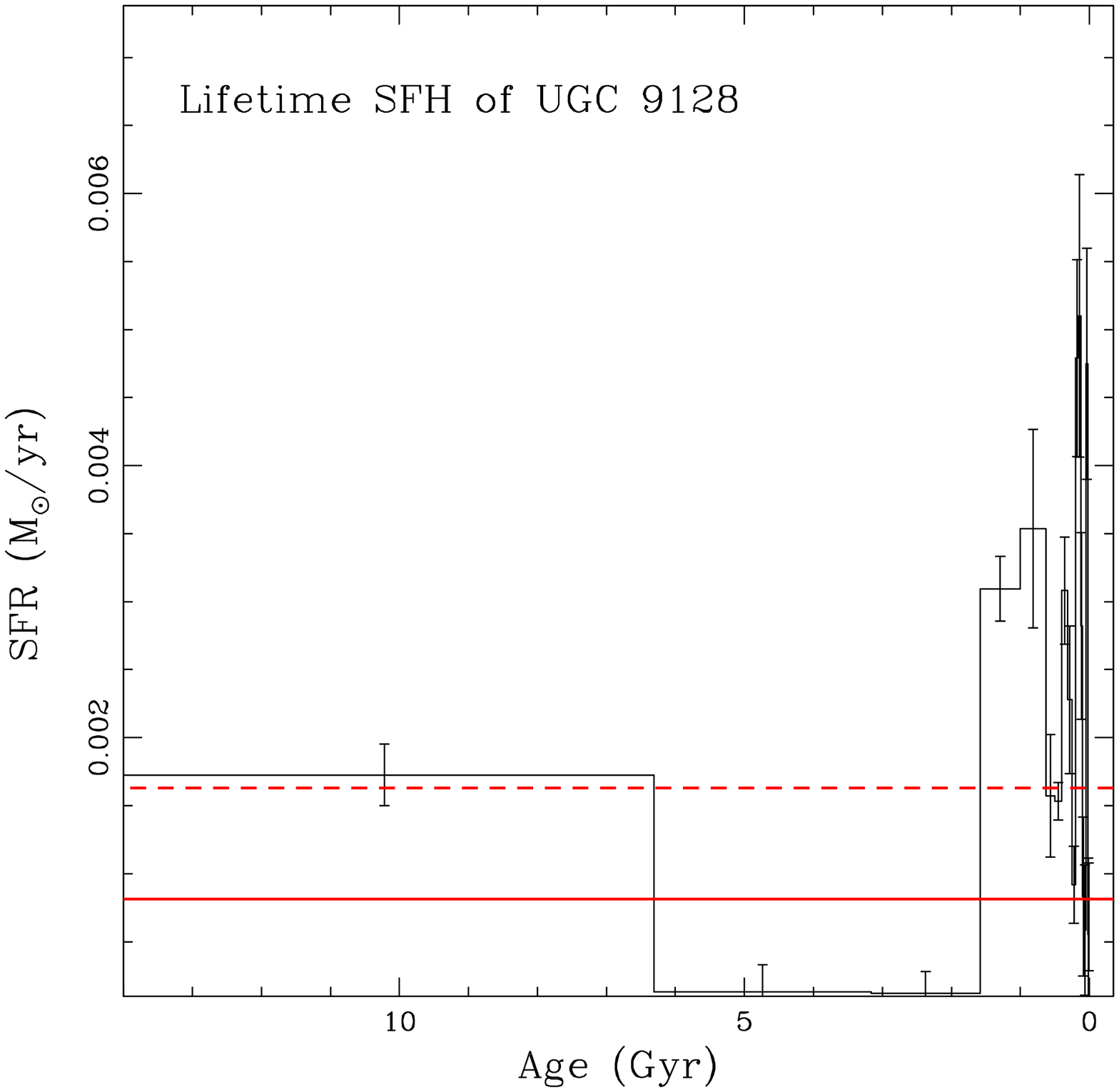,width=0.5\linewidth,clip=} &
\epsfig{file=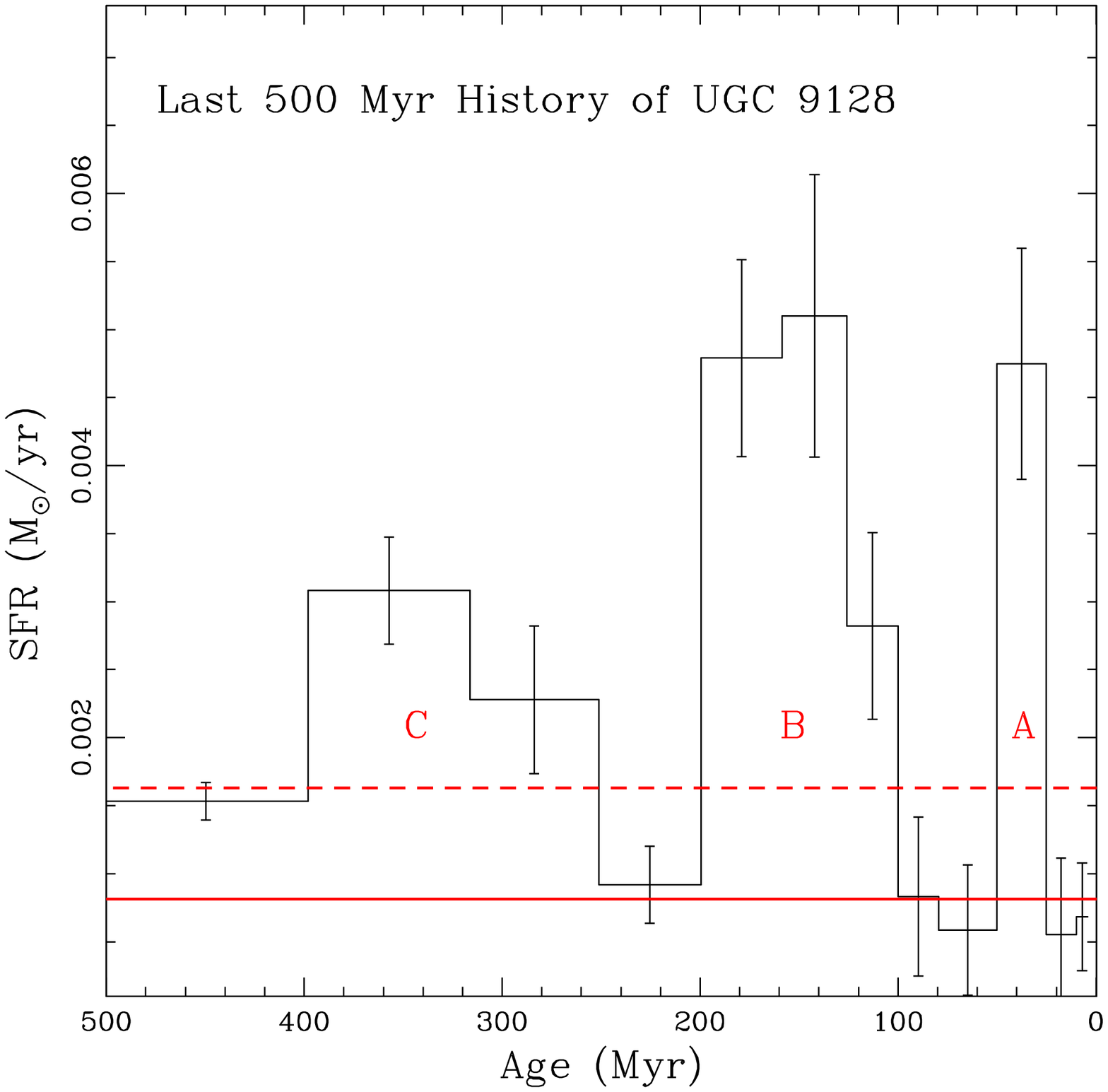,width=0.5\linewidth,clip=}
\end{tabular}
\caption{\scriptsize{This four panel figure shows the image, photometry, and SFH results for UGC~9128. Panel (a) presents the V band image of UGC~9128 from the HST ACS instrument. Panel (b) shows the CMD with average photometric errors per magnitude bin. The MS and HeB stages of evolution are marked. The regions on the CMD labeled (A), (B), and (C) correspond to three different episodes of recent SF that are not spatially coincident in the galaxy. Event (A) in the CMD includes BHeB stars, RHeB stars, and MS stars. Events (B )and (C) include the BHeB stars highlighted in CMD. The RHeB stars from these events are blended with the red clump while the MS stars born simultaneously with the BHeB stars in regions (B) and (C) lie below the photometric depth in the CMD. The reconstructed SFH for the lifetime of UGC~9128 is presented in panel (c). The solid red line indicates the average SFR over the past 6 Gyr (i.e., b$_{recent}=1$) and the dashed red line is twice this average (i.e., b$_{\mathrm{recent}}=2$). An expansion of the SFH during the last 500 Myr is presented in panel (d). The SF events (A), (B), and (C) identified in the CMD in panel (b) and the intervening periods of quiescent SF are reconstructed in the SFH.}}
\label{fig:ugc9128}
\end{figure}

We reconstructed the past rate of star formation (SFR$(t)$) for each galaxy by applying a CMD fitting technique \citep{Dolphin2002} with stellar evolutionary models from \citet{Marigo2007} to the stellar photometry. During fitting, the distance and extinction were allowed to vary within limited ranges, and the metallicity was constrained to increase with time except in three cases (Antlia, NGC~2366, and Holmberg II). The metallicity in these three systems could be fully constrained by the information contained in the deeper photometry reaching a magnitude below the red clump. The lifetime SFH results of the fitting program (i.e., SFR($t,Z$)) are presented in Paper I for eighteen galaxies and in \citet{Weisz2008} for two galaxies (DDO~165 and Ho~II). As a representative example of our results, we present the complete SFH of UGC~9128 in the bottom left panel of Figure~\ref{fig:ugc9128}. 

UGC~9128 provides an interesting example of how the features in a CMD translate into a SFR (see Figure~\ref{fig:ugc9128}). The distribution of stars along the helium burning (HeB) sequences is clumpy, suggesting that the SF has been elevated during several distinct episodes interspersed with lower levels of SF. We have highlighted stars produced during three periods of elevated SF labeled as events (A), (B), and (C) in the CMD. The most recent star formation event (A) is traced by BHeB stars, RHeB stars, and MS stars in the CMD. The earlier events (B) and (C) are traced by BHeB stars, as well as by RHeB stars that are blending into the red clump and MS stars that fall below our photometric depth. These three distinct SF events are recovered in the last 500 Myr in the SFH as shown in the final panel of Figure~\ref{fig:ugc9128}. Event (C) contains the lowest luminosity BHeB stars in the three events corresponding to the lowest mass (i.e., oldest aged) BHeB stars. This SF event is seen in the SFH as an elevated SFR from $\sim250$ Myr ago to $\sim400$ Myr ago. More recently (i.e., above the luminosity of event (C) in the CMD) is a period of relative quiescent SF followed by another increase in SFR (the clump of BHeB stars labeled event (B)). This more recent SF event is recovered in the SFH beginning $\sim200$ Myr ago and lasting $\sim100$ Myr. Similarly, event (B) was followed by a comparative lull in SF before experiencing another increase in SF from $\sim50$ Myr to $\sim25$ Myr ago, event (A). UGC~9128's unique SFH demonstrates how directly the HeB stars correlate to SFRs at recent times. 

In Figure~\ref{fig:sfh}, we show the SFH over the last 1.5 Gyr for all twenty galaxies. These recent SFHs share the common characteristic of elevated SF at recent times, as expected for our sample selection. However, there are differences within the sample. For example, some galaxies show a sustained, elevated rate of SF over a few hundred Myr (e.g., ESO~154$-$023, NGC~4068, UGC~4483). Others have significant fluctuations in their SFRs on shorter timescales  (e.g., UGC~9128, NGC~4214). Importantly, five galaxies show ``fossil" bursts, for which the SFR has returned to a lower level in the most recent time bins (e.g., Antlia, UGC~9128, NGC~4163, NGC~625, and NGC~6789). One galaxy, NGC~6822, shows a burst just beginning with elevated levels of SF only in the last 50 Myr.

\subsection{Photometric Depth and Temporal Resolution of SFHs}
As the photometric depth of each galaxy reaches below $\sim2$ mag below the TRGB, the recent SFHs in our sample have uniform temporal resolution regardless of the exact photometric depth. As demonstrated in the case of UGC~9128 in Figure~\ref{fig:ugc9128}, the HeB stars can be unambiguously age-dated from an optical CMD due to the rapid luminosity and spectral evolution of the youngest, most massive stars, all of which are well-sampled in our CMDs. Thus, the recent SFHs of starburst galaxies are not only robustly derived, but temporal resolution ranging from $\delta$t$\sim10$ Myr to a few 100 Myr, depending on look-back time, can also be achieved universally at recent times (t$\ltsimeq$1 Gyr).

The time resolution of the intermediate and ancient SFH is, in part, defined by the depth of photometry.  Between 1$<$t$<$2 Gyr, the SFRs are anchored by stars in the RGB and AGB phase of evolution above and just below the TRGB that have been shown to correlate to SF in this time period \citep[e.g.,][]{Mateo1998, Gallart2005}. At intermediate times, the derivation of the SFRs is complicated by the age-degeneracy of RGB stars. However, the SFH can be still constrained by bright AGB stars that lie above the TRGB as these stars do not suffer from the same age-degeneracy as the RGB stars, providing the photometry reaches deeper than $\sim2$ mag below the TRGB \citep[][see Fig. 4]{Gallart2005, Weisz2008}. At ancient times, deeper photometry reaching depths of the oldest MS turn-off (oMSTO) is needed to break the age-degeneracy of stars in the CMD and hence the co-variance of the SFR between time bins. Thus, without deep photometry, larger time bins are justified at higher look-back times.

As the time resolution at older look back times depends on the depth of photometry, we measured the inherent time resolution of our data by reconstructing the SFH of synthetic photometry with depths corresponding to our deepest (I$\sim2$ mag), typical (I$\sim-1$ mag) and shallowest (I$\sim-2$) data. For each photometric case, we input SF that occurred over a short time frame (i.e., log($\delta$t)$=$0.05) at different look-back times. The results from the artificial star tests for our real data were used in the SFH derivation thereby taking into account completeness and crowding effects. We present the results of these tests in Figure~\ref{fig:timeres}. The top panel shows the SF input at 100 Myr, 200 Myr, 450 Myr, and 815 Myr as dashed lines for three different photometric depths; the recovered normalized SFRs are shown in blue (100 Myr), red (200 Myr), cyan (450 Myr), and magenta (815 Myr). Generally, these results show that the SF is well-recovered with the peak SF occurring in the correct time bin with some SF spreading to the adjacent time bins. This ``spread'' defines the inherent time resolution of the data. For comparison, we overlaid the time binning we employ in our SFH results on these plots as shaded gray regions. The time resolution of the SFH is well-suited to measuring how long SF events last at recent times.

\section {Durations of Starbursts in Dwarf  Galaxies \label{durations}}

\subsection{Quantifying Starbursts}
The measurement of the duration of a starburst event requires a definition of ``bursting''. To establish a quantitative threshold, we apply a metric to evaluate the recent SFRs relative to the historical SFRs over the past 6 Gyr within a host galaxy, as discussed in \citet{McQuinn2009}. We used a birthrate parameter (b$={\mathrm{SFR}}~/<{\mathrm{SFR}}>$) as defined by \citet{Scalo1986} which compares the current SFR with the lifetime average SFR. A normalized SFR removes the bias against identifying bursting SFRs in smaller galaxies where the absolute SFR is generally lower than in larger galaxies. However, rather than using the lifetime average SFR as a baseline, we used the average SFR over the last few Gyr, providing a better comparison of the recent SFRs with the more recent evolutionary state of the host galaxy. This choice both avoids suppressing b by excluding the higher levels of SF typically expected during the initial period of galaxy assembly and leverages the most secure part of the SFH. We use our modified birthrate parameter (b$_{\mathrm{recent}}=$SFR~/$<$SFR$>_{\mathrm{0-6~Gyr}}$) to define a threshold for bursting SFRs; a value of b$_{\mathrm{recent}}>$ 2 identifies a burst \citep{Kennicutt2005} and b$_{\mathrm{recent}}\simeq1$ marks the beginning and end of a burst event. In Figure~\ref{fig:sfh}, we show the average SFR from the last 6 Gyr (b$_{\mathrm{recent}}=1$) as a solid red line and twice this average (b$_{\mathrm{recent}}=2$) as a dashed red line.

The duration of the starburst event can be measured directly as the amount of time b$_{\mathrm{recent}}\geq1$ once the b$_{\mathrm{recent}}>2$ threshold has been reached. The time measurement extends from the midpoints of the beginning and end time bins. The uncertainties on the durations are the width of the longest time bin contributing to the duration. The resulting durations of the burst events are shaded in blue in Figure~\ref{fig:sfh}. Fifteen galaxies have starburst durations between $450\pm50$ and $640\pm190$ Myr as listed in Table~\ref{tab:durations}.\footnote{Note: We measure slightly different durations for NGC~4163, NGC~4068, and IC~4662 than previously reported in \citet{McQuinn2009}. As discussed in Paper I, we constrained the metallicity of the galaxies to increase with time, which gives improved constraints on the recent SFH. As a result, the SFH of these three systems changed, altering the measured duration of the starbursts by 25\% for NGC~4068 and IC~4622 assuming a double burst, and 50\% for NGC~4163.} The longest bursts we identify are in UGC~9128 and DDO~165, which show bursts lasting $1300\pm290$. The average burst duration is $\sim600$ Myr. In fifteen cases, the SFR remains above the 6 Gyr historical average in the most recent time bin, indicating that the starburst is ongoing. The durations for these starburst events are thus lower limits. The average SFRs over the past 6 Gyr, peak SFRs and corresponding b$_{\mathrm{recent}}$ value during each burst event are also listed in Table~\ref{tab:durations}.
 
Note that the duration measurement is affected by the inherent time binning of the SFH. The temporal resolution over the last $100$ Myr is fine enough that we can discern and measure fluctuations in the SFR over short timescales and pinpoint the value of b$_{\mathrm{recent}}$ every $20-50$ Myr or so. In contrast, between $650-1000$ Myr ago, the inherent time resolution of the data reduces to $350$ Myr. One can imagine a burst beginning in the latter half of this time bin but the b$_{\mathrm{recent}}$ value is suppressed below unity due to averaging with a lower SFR earlier the same time bin. In such a case, this entire time bin would not be considered part of the burst event and the measured duration would be underestimated. The reverse is also possible. The resolution of time bins has a smaller effect on the final duration measurement for bursts contained in the most recent $650$ Myr due to the finer temporal resolution. This effect is quantified in the uncertainties listed for the durations which increase with larger time binning.

One concern is that our results are in some way dependent on how we bin the oldest SF. In order to investigate the effect on our results of dividing the intermediate and ancient SF into separate bins, we calculated the duration of the starbursts without distinguishing between intermediate and ancient time bins (i.e., the classical b parameter $\equiv$ SFR / $<$SFR$>_{lifetime}$). Given that Local Group galaxies, whose photometry reaches below the red clump, show detectable SF older than 10 Gyr based on red giant population and/or RR Lyr stars \citep{Mateo1998}, we assume the lifetime of the galaxies extend back to our oldest time bin. In most cases, including the ancient SF raises the baseline average SFR due to the higher rates of SF typical during the initial galaxy assembly period in the early universe. Thus, with a higher baseline SFR average, a higher rate of SF is needed to reach the bursting threshold of b$>$2. Nonetheless, $\textit{only}$ three of the galaxies in our sample (Antlia, NGC~4163, NGC~784) did not meet this criterion and would not be considered bursts by this metric. The remaining seventeen galaxies are still identified as starbursts. Perhaps surprisingly, the duration of the bursts measured with this alternate metric are shortened in $\textit{only}$ six of the seventeen systems, lowering the average duration of the sample from $\sim$600 Myr to $\sim$450 Myr. Thus, our main scientific conclusion that starbursts have longer durations is robust and the uncertain splitting of the time bins is of negligible import.

\subsection{Cases of Complex SFHs}
Our criterion for defining a starburst serve as a prescriptive guide for measuring a burst duration but we stress that the criterion is a guide and the distinction between ``regular'' and ``bursting'' SF is subjective in some cases. In thirteen of the galaxies (e.g., ESO~$154-023$, NGC~784, NGC~5253, among others), the SFR profile increases substantially from one time bin to another and the duration measurement is easily made from analyzing the b$_{\mathrm{recent}}$ parameter. In NGC~6822, the burst began $\sim 50$ Myr ago. We measure the lower limit of the duration in this system to be $20\pm13$ Myr. Given the much longer durations of the other galaxies, this burst is likely just beginning. In six systems the SFH is more complex and thus we discuss each system in turn. 

For NGC~2366 and Ho~II, the burst is preceded by time bins whose b$_{\mathrm{recent}}$ is equivalent to unity within the measured uncertainties (we refer the reader again to the blue shaded regions and red lines in Figure~\ref{fig:sfh}). By our definition, these time bins should be included in the duration measurement. Yet, these SFRs continue for hundreds of Myr at roughly the historical average SFR. The overall SFR profile suggests that the bursts had not yet begun in these time bins. The lower temporal resolution at these older times also affect the b$_{\mathrm{recent}}$ calculation. We chose to exclude these time bins from the duration measurement. 

In the third case, NGC~6789, the b$_{\mathrm{recent}}$ index drops just below unity approximately $350$ Myr ago after which the SFR increases again pushing the b$_{\mathrm{recent}}$ close to 2. There is some co-variance between adjacent time bins in the SFHs and this short dip in SFR surrounded by higher levels of SF suggests that the burst is continuing. The SFR profile indicates that the burst event, while weaker than others in our sample, has continued to elevate the SF above the average until our most recent time bins. This conclusion is supported by looking at the overall SFR profile over the past 1.5 Gyr. The period preceding the burst is a quiescent period experiencing very little SF. We chose to include the SF $\sim350$ Myr ago as part of the burst event which extends the burst duration measurements from $600$ Myr ago until the last $\sim100$ Myr. The profile is similar to UGC~6456 where the burst event is clearly defined and extended. 

In the fourth case, UGC~4483, the SFR profile experiences a dip $\sim450$ Myr ago and the b$_{\mathrm{recent}}$ index is equivalent to unity within the measured uncertainties. The time bins on either side are elevated reaching b$_{\mathrm{recent}}$ values of 13 and 3.9. This suggests that the burst has continued although the system experienced a period of lower SF during the larger burst event. Likewise, in SBS~1415$+437$, the SFR dips to zero during one time bin flanked on one side with b$_{\mathrm{recent}}$ values of 5.7 and 6.3. We chose to define the burst event inclusive of this period of quiescence based on the overall SFR profile. Note that the photometry of SBS~$1415+437$ falls $\sim 0.5$ mag short of our photometric depth requirement, thus the ancient SFH is not well-constrained. However, the recent SFH is well-constrained and the SFR profile in the last 500 Myr clearly defines a burst. We measure the duration of the starburst in SBS~$1415+437$ despite the uncertainties in the past SFR average and include it for comparison with others in our sample.

Finally, the well$-$studied starburst galaxy, NGC~625 \citep{Cannon2003}, does not fit our metric for identifying a starburst. The SFH of NGC~625 is similar to that of Leo~A in that the initial SF is suppressed \citep{Cole2007} until $\sim 3$ Gyr ago. The photometry used to derive the SFH of NGC~625 reaches nearly three magnitudes below the TRGB giving confidence to the low SFRs derived for the 6$-$14 Gyr time bin. The result of the later onset of SF is that the average SFR over the past 6 Gyr is high and the elevated levels of SF in the last 500 Myr fall below our starburst threshold. Although the unique SFH of NGC~625 does not fit our metric, evidence at other wavelengths classify this system as a starburst reinforcing the premise that the identification of starburst (i.e., the difference between ``regular'' and ``bursting'' SF) is not always clear. For example, if we apply a classical birthrate parameter utilizing the SFR average over the lifetime of the galaxy, the recent SFRs do meet the starburst threshold. Using the elevated SFRs in the recent history of NGC~625 to guide the measurement, the duration of this likely starburst event is $450\pm50$ Myr as seen in Figure~\ref{fig:sfh}, much like the majority of others measured in this study. We include it in our duration measurements for comparison. Excluding this one unusual system would not change our scientific conclusions.

\section{Bursts: SF on Timescales $>$ Few 100 Myr\label{bursts}}
The durations of the burst events fall unambiguously on the scale of hundreds of Myr. The bursts are sufficiently robust that they are not shut off by an individual flicker of SF. Moreover, the longer durations suggest that a starburst event is a galaxy-wide phenomenon, which likely includes SF in both clusters and in the field regions of the galaxies \citep{McQuinn2009}. Paper III will specifically investigate the spatial distribution and possible migration of SF in the stellar clusters and diffuse regions within the starbursts to better characterize the nature of starbursts.

The burst durations can be compared with the characteristic timescale to communicate an event or disturbance across a galaxy. This dynamical timescale can be estimated as the rotation period for dwarf galaxies as these systems exhibit solid$-$body rotation \citep{Skillman1988} (i.e., $t_{dyn}\simeq2\pi Re/V$ where $R_{e}$ is the effective radius of the galaxy and $V$ is the rotational velocity). We have calculated the dynamical timescale for eighteen\footnote{Rotational velocity data were not available for SBS~1415$+437$ and NGC~6789.} of the galaxies using the major axis radii of the galaxies and the inclination corrected maximum rotation curve velocities. The radii and velocity measurements were obtained from the HyperLeda database \citep{Paturel2003} with the exception of NGC~6822 whose maximum rotational velocity was obtained from \citet{Weldrake2003}. The dynamical times calculated are listed in Table~\ref{tab:durations} and are compared with the starburst duration we measured for each galaxy. The ratio of the duration to dynamical timescales for fifteen starbursts range from $0.70-4.6$ with a mean ratio of 2.6. Galaxy UGC~9128, with a duration of over 1 Gyr, has a larger duration/dynamical timescale ratio of 10; NGC~6822 has ratio of 0.14 as the burst is just beginning in this system. The dynamical timescales are plotted as green horizontal bars in the SFHs in Figure~\ref{fig:sfh} and can be compared to individual galaxy duration measurements. Previous authors have proposed that if starbursts persist only on the characteristic timescale of a system or shorter, then they would likely be self-extinguishing events destroyed by their own energy output \citep{Meurer2000}. Durations comparable to or longer than the dynamical timescales support our conclusion that these events are not self-quenching on the timescale of massive star lives as discussed in Paper I. 

The starburst durations we measure are clustered around $450$ Myr, as shown in Figure~\ref{fig:histo_durations}. It is important to note that 75\% of these systems have not yet finished; we are measuring lower limits for the burst durations in these systems. The peak of the distribution in Figure~\ref{fig:histo_durations} would likely shift to longer timescales if we had a sample of exclusively fossil bursts. The distribution of the dynamical timescale is plotted as a dashed red histogram in Figure~\ref{fig:histo_durations}. This distribution is similar to the distribution of the burst durations but is shifted towards shorter times suggesting the burst process is global in nature. Rapid self-quenching of SF would limit starburst durations to $\sim10$ Myr, indicated by the dashed blue line.

\section{``Flickering": SF on $\sim$10 Myr Timescales\label{flickering}}
Burst durations on the scale of hundreds of Myr differ substantially from the often reported timescale of $5-10$ Myr (see references in \S\ref{intro}). The nearly two orders of magnitude difference between these two scales is the result of measuring two different scales of SF events; i.e., whether one is measuring the duration of a complete starburst event in a galaxy or measuring the duration of a pocket of enhanced SF made up of one or a few stellar clusters or associations within a galaxy. Both scales of SF are evident in our SFHs. We have highlighted the burst events in blue in Figure~\ref{fig:sfh}. Now we draw the reader's attention to the most recent time bins with the finest resolution (t $\sim 4-10$ Myr, t $\sim 10-25$ Myr) presented in Figure~\ref{fig:sfh}. Over these short timescales, there are rapid variations in the SFRs seen in many of the galaxies corresponding to local ``flickering'' of SF on small scales. This flickering occurs within a larger envelope of enhanced SF on timescales greater than a few 100 Myr. The larger starbursts events are likely comprised of numerous individual pockets of flickering SF in stellar clusters or associations as well as SF in diffuse field regions and the starburst duration is the summation of all these smaller scale events. 

Obviously, small scale increases in the SFR can occur outside of a burst event. In other words, flickering SF and the formation of stellar clusters and associations do not occur solely in starbursts. For example, consider the SFH of the Antlia dwarf galaxy in Figure~\ref{fig:sfh}. The burst event in Antlia ended nearly 200 Myr ago. Yet, we see elevated SF during one additional time period (t $\sim 10-25$ Myr ago) where the b$_{\mathrm{recent}}$ index exceeds 2. This increase in SF is temporally isolated from the larger burst event and is likely caused by SF in an individual stellar association or small cluster. We do not include this flickering SF as part of the burst event as the surrounding time bins have lower levels of SF. One of the implications of longer burst durations is that the burst is likely occurring on large spatial scales within a galaxy and small increases in SFR are qualitatively different than bursting SF. We will explore this in detail in Paper III with a spatial analysis of the SF. Note, we are only sensitive to this episode in Antlia because of the fine time resolution in the SFH 25 Myr ago. If this isolated flickering event had occurred 200 Myr ago, it would have blended into the SFR derived for the larger time bin.

The variations of SFR on short timescales ($\delta$t $\sim 5$ Myr) are also seen when the SFR results in our most recent time bin of $4-10$ Myr are compared with another starburst indicator, H$\alpha$ emission, as discussed in Paper I. The timescales for our most recent time bin and H$\alpha$ emission overlap but do not completely coincide. As one would expect, the results of these two indicators agree in most galaxies in the sample (i.e., where there is significant H$\alpha$ emission, we find b$_{\mathrm{recent}}$  values above two in the $4-10$ Myr time bin and vice versa). Yet in four galaxies, the two indicators disagree suggesting that the SF activity has changed rapidly on timescales of a few Myr.  This short timescale is the same as what is expected for the evolution of massive stars. 

The characteristic timescale to communicate an event or disturbance for ``flickering'' SF is the crossing time for a typical star cluster ($t_{cross}=2R_{e}/\sigma$ where $\sigma$ is the velocity dispersion of the stars). Assuming a typical radius of a star cluster to be a few pc with a velocity dispersion of $\simeq10$ km s$^{-1}$ \citep[e.g.,][]{Smith2000, Larsen2008}, the crossing time is $\simeq1$ Myr. This value agrees with the typical dynamical star cluster timescale of a few Myr reported by \citet{Heckman2005} using a timescales defined by $t_{dyn}\simeq(G\rho)^{-1/2}$. These cluster crossing times support the conclusion that the short durations reported for bursts are measuring the SF phenomenon within star clusters, associations, or some equivalently smaller region of SF. 

The amplitude of the flickering can be used to estimate the stellar mass created in these smaller scale SF events. Using a range of SFRs from $0.001-0.1$~\msun\ yr$^{-1}$ typical of the sample dwarf galaxies and a timescale of 10 Myr, the total stellar mass produced ranges from $10^{4}-10^{6}$~\msun, which are typical masses of stellar associations and clusters. These numbers can be compared to the larger values of $10^{7}-10^{9}$~\msun\ of stellar mass created during the bursts events of these same galaxies.

The differences in timescales, dynamical times, and amount of stellar mass produced highlight the different physical processes governing local, small scale SF and large scale burst events. The flickering SF is governed by local physical conditions (e.g., gas density, feedback from massive stars, etc.) and is likely to be temporarily suppressed/quenched locally \citep{Omukai1999, Nishi2000}.  The same feedback from massive stars that may quench the flickering SF does not disrupt/quench the overall burst event in our sample of galaxies. Thus it appears the non-equilibrium energy output and mass transfer characteristic of stellar winds and supernovae that have been proposed to quench SF  \citep{Ferguson1998, Thornley2000, Stinson2007, Kaviraj2007} cannot be assumed universally on galactic scales.

\section{Stellar Mass and Energy Created in a Burst \label{mass_energy}}
Longer duration bursts have a greater impact on the evolution of the host galaxy than short bursts. Two interesting parameters to quantify are the percent of stellar mass created during a burst and the energy produced by the burst, which can potentially be deposited into the ISM. We have calculated the percent of stellar mass created in each burst relative to the stellar mass created over the lifetime of each galaxy directly from the SFHs. The results range from 3\%$-$26\% of the mass being created during the burst; the individual numbers for each galaxy are listed in Table~\ref{tab:mass_energy}. The percentage is an upper limit for SBS1415$+437$ whose ancient SFH remains uncertain due to the limited depth of photometry. The galaxies showing bursts with lower absolute SFRs (i.e., Antlia) or SFRs exceeding the past 6 Gyr average in only one or two time bins (i.e., UGC~6456, NGC~4163, and NGC~6789) lie at the lower end of this range. Galaxies with longer durations and stronger bursts, such as UGC~9128, DDO~165, NGC~4214, lie at the upper end of this range. Additionally, we compare the total amount of stellar mass calculated form the SFH with the luminosity of the galaxies from published absolute magnitudes in the B band, adjusted for extinction from Galactic dust maps \citep{Schlegel1998}. The mass derived from the SFH does not incorporate all the mass from the optical extent of the galaxies as the field of view in some cases does not extend to the outer optical limits. While this technically makes the mass a lower limit, it is a small effect as most of the optical area is covered in each system and the amount of mass at these outer radii is small compared to the mass in the inner parts of the systems. The one exception is NGC~6822 for which the observations do not cover the main extent of the stellar populations. The mass$-$to$-$light ratios ($M_*$/$L_B$), in solar units, range from $1.0\pm0.2$ to $8.0\pm1.0$. The highest ratio is for Antlia dwarf galaxy, the lowest surface brightness galaxy with an absolute magnitude of $-10.14$. These $M_*$/$L_B$ ratios are consistent with the expected ratios for SF galaxies thus providing an independent check on the derived SFHs.

We calculated the energy produced in each burst using our SFHs and STARBURST99 \citep{Leitherer1999, Vazquez2005}. STARBURST99 is an evolutionary synthesis code that uses a known SFR and stellar evolution models to predict spectrophotometric properties, the energy distribution, and other related properties of galaxies with active star formation. Of particular interest here is the ability of STARBURST99 to simulate the energy produced by stellar winds and supernovae at a given SFR. Using an instantaneous burst of SF with a fixed mass of $10^{6}$~\msun\, coupled with the same IMF and stellar evolution models employed in the SFH code as done in \citet{Weisz2009}, we simulated the energy output of SF for low metallicity galaxies with a time resolution of 5 Myr. This energy output for a $10^{6}$~\msun\ instantaneous burst was scaled by the stellar mass created every 5 Myr for the actual observed bursts and integrated over the duration of the burst. The total energy for the bursts ranges from $10^{53.9} - 10^{57.2}$ ergs; individual values are listed in the final column of Table~\ref{tab:mass_energy}. There is a clear trend that higher luminosity galaxies have a higher amount of energy produced in a burst. The significant amount of energy deposited into the ISM, particularly in the cases where the energy is $>10^{56}$ ergs, will have an important impact on the gas dynamics and future SF activity of the host galaxy. We will study the spatial distribution of the SF activity in Paper III and explore whether this energy is deposited in concentrated areas or over a significant fraction of the galaxy. 

\section{Conclusions \label{conclusions} }
The starburst events in the the twenty dwarf galaxies studied present bursts lasting on the order of hundreds of Myr. The shortest durations measured were $\sim450$ Myr, the longest were 1.3 Gyr. The majority of the galaxies display bursts with durations ranging from $\sim450$ Myr to $\sim650$ Myr but many are lower limits as the SFR is still elevated at the most recent times. The inherent time resolution of our data allows us to accurately derive the duration of starbursts occurring over the past $\sim1$ Gyr. The ratio of the durations to the dynamical timescales of the galaxies ranges from $0.70 - 4.6$ for sixteen of the galaxies with one outlier, UGC~9128, having a ratio of duration to dynamical time of 10. These longer duration times rule out the theory that these starbursts are ``quenched'' by the feedback processes of the massive stellar populations created. Shorter durations often cited on the order of $5-10$ Myr are measuring small pockets of enhanced SF or ``flickering'' that contribute to a burst event but do not represent the extant of a burst.  The bursts deposit $10^{53.9}-10^{57.2}$ ergs of energy into the interstellar medium through stellar winds and supernovae and produce 3\%$-$26\% of the host galaxy's mass.

\section{Acknowledgments}
Support for this work was provided by NASA through grants AR-10945 and AR-11281 from the Space Telescope Science Institute, which is operated by Aura, Inc., under NASA contract NAS5-26555. E.~D.~S. is grateful for partial support from the University of Minnesota. J.~J.~ D.  was partially supported as a Wyckoff fellow. K.~B.~W.~M. gratefully acknowledges Matthew, Cole, and Carling for their support. This research made use of NASA's Astrophysical Data System and the NASA/IPAC Extragalactic Database (NED) which is operated by the Jet Propulsion Laboratory, California Institute of Technology, under contract with the National Aeronautics and Space Administration. Finally, we would like to acknowledge the usage of the HyperLeda database (http://leda.univ-lyon1.fr).

{\it Facilities:} \facility{Hubble Space Telescope}


\begin{deluxetable}{llllr@{$\pm$}lll}
\tabletypesize{\scriptsize}
\tablewidth{0pt}
\tablecaption{SFRs, Burst Durations, and Timescales \label{tab:durations}}
\tablecolumns{8}
\tablehead{
\colhead{}				&
\colhead{$<$SFR$>_{\textrm{6-0 Gyr}}$}	&
\colhead{Peak b$_{\mathrm{recent}}$}		&
\colhead{Peak SFR of Burst}		&
\multicolumn{2}{c}{Duration}		&
\colhead{$\tau_{dyn}$}			&
\colhead{}				\\
\colhead{Galaxy}			&
\colhead{(10$^{-3}~$\msun\ yr$^{-1}$)}	&
\colhead{of Burst}			&
\colhead{(10$^{-3}~$\msun\ yr$^{-1}$)}	&
\multicolumn{2}{c}{(Myr)}		&
\colhead{(Myr)}				&
\colhead{Dur/$\tau_{dyn}$}		\\
\colhead{(1)}				&
\colhead{(2)}				&
\colhead{(3)}				&
\colhead{(4)}				&
\multicolumn{2}{c}{(5)}			&
\colhead{(6)}				&
\colhead{(7)}				
}
\startdata
Antlia Dwarf	& $0.15\pm 0.01$	& $3.5  \pm0.9$ & $0.52	\pm0.01$&    $~640$&190     & 190$\pm$20	&       3.3$\pm$1.0	\\
UGC 9128	& $0.82\pm 0.07$	& $6.3  \pm1.4$ & $5.1	\pm1.0$ &    $~1300$&300    & 120$\pm$10	&       10.$\pm$7	\\
UGC 4483	& $0.79\pm 0.06$	& $14 	\pm3$ 	& $11	\pm2$ 	&    $>810$&190  & 200$\pm$23	&       4.1$\pm$1.0 	\\
NGC 4163	& $4.0 \pm 0.1$		& $2.9  \pm0.6$ & $12	\pm3$ 	&    $~460$&70      & 240$\pm$20	&       1.9$\pm$0.1	\\
UGC 6456	& $3.0 \pm 0.2 $	& $7.6  \pm1.1$ & $23	\pm3$ 	&    $>570$&60      & 160$\pm$30	&       3.5$\pm$0.7	\\
NGC 6789	& $3.9 \pm 0.1$		& $3.8  \pm1.3$ & $15	\pm5$	&    $~480$&70      &  ...	     	&       ...		\\
NGC 4068	& $10. \pm 0.3$		& $4.7  \pm0.3$ & $46	\pm3$ 	&    $>450$&50      & 270$\pm$20	&       1.7$\pm$0.1	\\
SBS 1415+437	& $12  \pm 2$		& $12 	\pm2$	& $150  \pm10$	&    $>450$&50      & ...		&       ...		\\
DDO 165	 	& $13  \pm 1$		& $6.5	\pm0.5$	& $80.	\pm5$ 	&    $>1300$&300    & 710$\pm40$	&      	1.8$\pm$0.4	\\
IC 4662		& $9.8 \pm 0.6$		& $7.7  \pm1.6$ & $76	\pm15$	&    $>450$&50      & 100 	    	&       4.6		\\
ESO 154-023	& $20. \pm 1$		& $6.4  \pm0.5$ & $120	\pm10$ 	&    $>450$&50      & 460	     	&       1.0		\\
NGC 2366	& $29  \pm 1$		& $5.6  \pm0.4$ & $160	\pm10$	&    $>450$&50      & 330		&       1.4	 	\\
NGC 625		& $29  \pm 1$		& $1.4  \pm0.1$ & $40.	\pm2$	&    $~450$&50	     & 720	     	&       1.6		\\
NGC 784		& $26  \pm 1$		& $4.5  \pm0.6$ & $120	\pm20$	&    $>450$&50      & 470	     	&       1.0		\\
Ho II	 	& $27  \pm 1$		& $6.5	\pm0.8$	& $180	\pm20$	&    $>570$&70	     & 820$\pm20$	&      	0.70$\pm$0.13	\\
NGC 5253	& $45  \pm 1$		& $9.0  \pm0.9$ & $400	\pm40$	&    $>450$&50      & 450$\pm$20	&       1.0$\pm$.0	\\
NGC 6822	& $2.3 \pm 0.1$		& $3.1  \pm1.1$ & $7.3	\pm2.5$	&    $>20.$&13	     & 140		&       0.14		\\
NGC 4214	& $43  \pm 2$		& $3.1  \pm0.9$ & $130  \pm40$	&    $>810$&190     & 400$\pm$10	&       2.0$\pm$0.5	\\
NGC 1569	& $21  \pm 2$		& $12	\pm1$	& $240	\pm10$	&    $>450$&50	     & 170$\pm$10	&       2.6$\pm0.3$ 	\\
NGC 4449	& $160 \pm 10$		& $6.0  \pm0.5$ & $970	\pm70$	&    $>450$&50      & 300$\pm$10	&       1.5$\pm$.1	\\
\enddata

\tablecomments{Col. (5) Durations are lower limits for ongoing bursts. Col. (6) Timescales were calculated using major axis measurements and maximum rotational velocity corrected for inclination from the HyperLeda database \citep{Paturel2003} with the exception of NGC~6822 whose maximum rotational velocity was obtained from \citet{Weldrake2003}. Uncertainties were not reported on the rotational velocities of five galaxies, thus the dynamical timescale lacks uncertainties in these cases.}

\end{deluxetable}


\begin{deluxetable}{lrrrrc}
\tabletypesize{\scriptsize}
\tablewidth{0pt}
\tablecaption{Stellar Mass and Energy Calculations \label{tab:mass_energy}}
\tablecolumns{6}
\tablehead{
\colhead{}				&
\colhead{M$_{B}$}			&
\colhead{Total Stellar}			&
\colhead{}				&
\colhead{\% Mass}			&
\colhead{log Energy of}			\\
\colhead{Galaxy}			&
\colhead{(mag)}				&
\colhead{Mass ($10^{6}$\msun)}		&
\colhead{$M_*$/$L_B$}			&
\colhead{Created in Burst}		&
\colhead{Burst (erg)}		
}

\startdata
Antlia Dwarf	& -10.14 & $7.3\pm0.9$	& $8.0\pm1.0$	& $3.2\pm0.5$	&   54.2\\
UGC 9128	& -12.45 & $19\pm3$ 	& $2.4\pm0.4$	& $24\pm5$	&   55.5\\
UGC 4483	& -12.68 & $15\pm3$	& $1.6\pm0.4$	& $26\pm7$	&   55.4\\
NGC 4163	& -13.75 & $140\pm40$	& $5.5\pm1.4$	& $2.8\pm0.8$	&   55.5\\
UGC 6456	& -13.85 & $68\pm27$	& $2.4\pm1.0$	& $5.1\pm2.2$	&   55.3\\
NGC 6789	& -14.60 & $100\pm30$	& $1.8\pm0.5$	& $3.9\pm1.4$	&   55.5\\
NGC 4068	& -14.96 & $320 \pm50$	& $4.1\pm0.7$	& $4.8\pm0.8$	&   56.0\\
SBS 1415+437	& -15.11 & $240\pm50$	& $2.7\pm0.6$	& $20.\pm4$	&   56.5\\
DDO 165	 	& -15.19 & $270 \pm50$	& $2.8\pm0.5$	& $19\pm4$	&   56.5\\
IC 4662		& -15.39 & $420 \pm90$	& $3.7\pm0.8$	& $5.2\pm1.2$	&   56.2\\
ESO 154-023	& -16.21 & $650\pm90$	& $2.7\pm0.4$	& $6.4\pm0.9$	&   56.5\\
NGC 2366	& -16.33 & $370 \pm50$	& $1.3\pm0.2$	& $12\pm2$	&   56.5\\
NGC 625		& -16.26 & $370\pm140$	& $1.5\pm0.6$	& $3.8\pm1.5$  	&   56.0\\
NGC 784		& -16.78 & $1100\pm100$	& $2.6\pm0.3$	& $4.7\pm0.7$	&   56.5\\
Ho II	 	& -16.92 & $450 \pm60$	& $1.0\pm0.1$	& $12\pm2$	&   56.5\\
NGC 5253	& -16.98 & $2200\pm300$	& $4.3\pm0.6$	& $4.7\pm0.7$	&   56.9\\
NGC 6822	& -17.86 & $28\pm4$	& ...		& $1.0\pm0.2$  	& 53.9\\
NGC 4214	& -17.02 & $400\pm100$	& $1.0\pm0.2$	& $18\pm5$	&   56.7\\
NGC 1569	& -17.76 & $1000\pm100$	& $1.0\pm0.1$	& $5.8\pm0.7$	&   56.6\\
NGC 4449	& -18.02 & $3000\pm500$	& $2.3\pm0.4$	& $7.0\pm1.5$	&   57.2\\
\enddata

\tablecomments{Col. (2) Integrated absolute magnitude of the galaxy adjusted for extinction in the B band. Cols. (3) The mass derived from the SFH does not incorporate all the mass from the optical extent of the galaxies as the field of view in some cases does not extend to the outer optical limits. While this technically makes the mass a lower limit, it is a small effect as most of the optical area is covered in each system and the amount of mass at these outer radii is small compared to the mass in the inner parts of the systems. The exception in NGC~6822 where the observations cover less than half of the optical disk. The total stellar mass is a lower limit for the galaxy. The mass-to-light ratio is not calculated for NGC~6822 as the mass and luminosity measurements cover significantly different areas of the galaxy.  Col. (4) Uncertainties are derived assuming errors are dominated by uncertainties in mass; uncertainties in luminosity are not included but are negligible at $\sim0.2$ mag. Col. (5) The percent of mass created in each burst are likely upper limits. The high SFRs in the galaxies prevent a complete census of the older (t$>6$ Gyr) populations. This is particularly true in the case of SBS1415$+437$ whose ancient SFH remains uncertain due to limited depth of photometry.}

\end{deluxetable}

\clearpage

\clearpage
\begin{figure}
\plotone{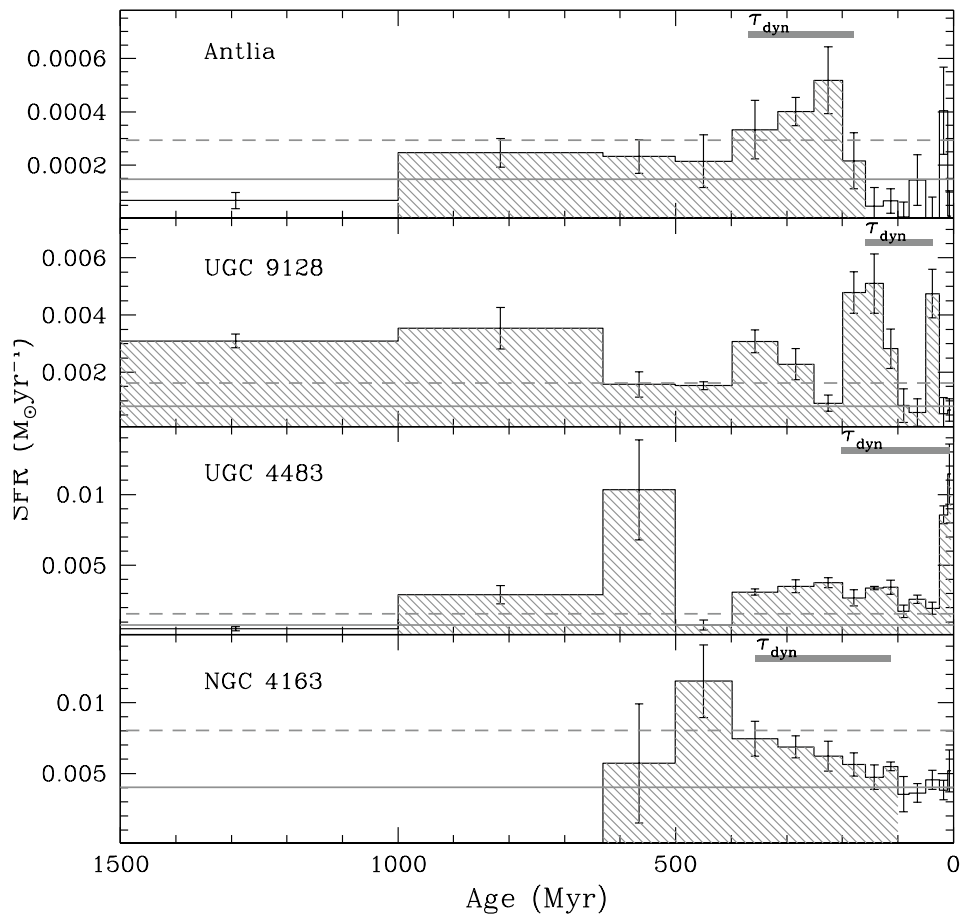}
\caption{The SFH of the last 1.5 Gyr for the twenty galaxies. The durations of the bursts are shaded in blue. The dynamical timescales of the galaxies are plotted in a green bar for comparison. The solid red lines indicate the average SFR over the last 6 Gyr for the galaxy (i.e., b$_{\mathrm{recent}}=1$) and the dashed red line is twice this average (i.e., b$_{\mathrm{recent}}=2$).}
\label{fig:sfh}
\end{figure}

\clearpage
\begin{figure}
\figurenum{\ref{fig:sfh}}
\plotone{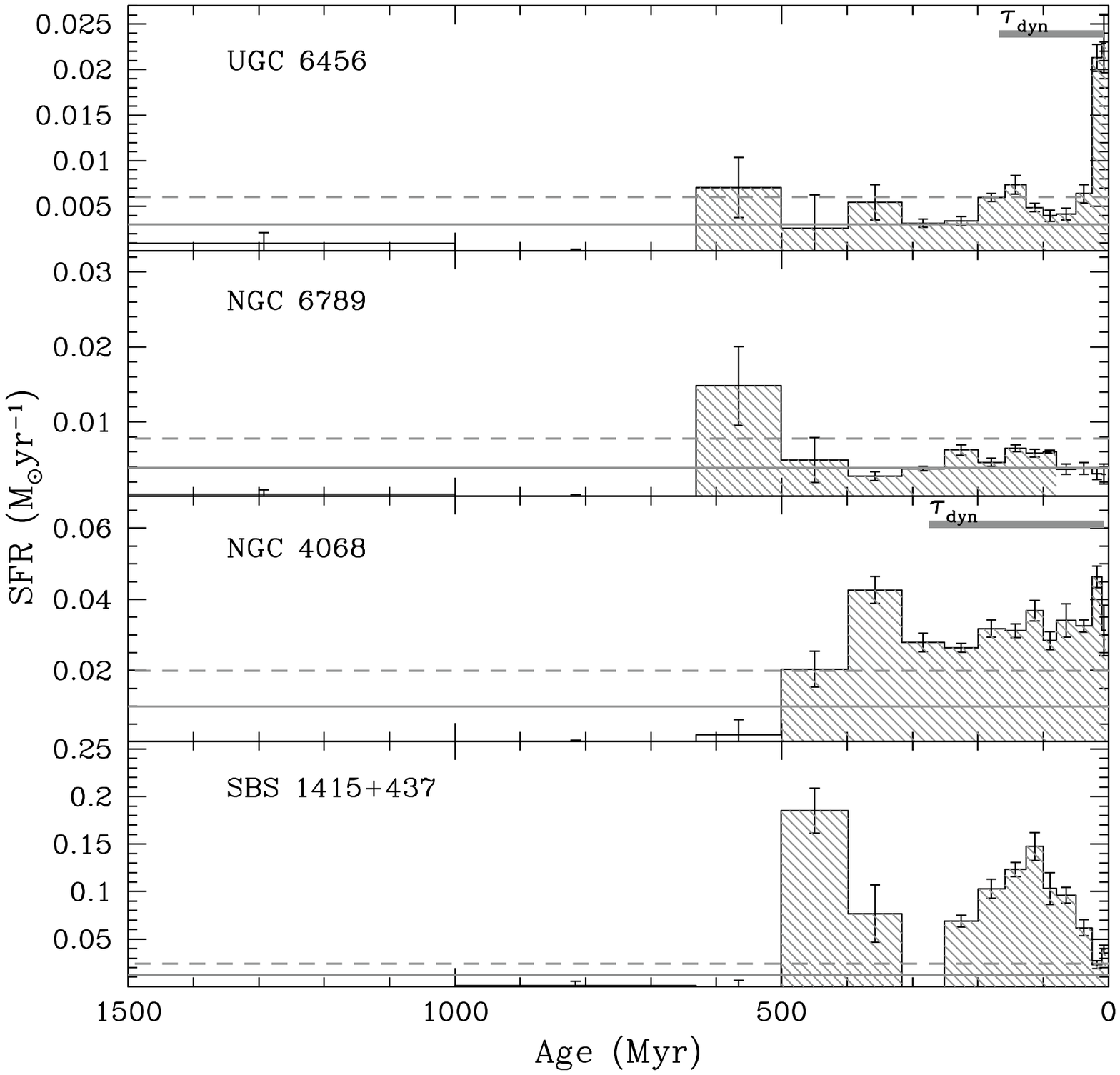}
\caption{\textit{Continued}}
\end{figure}

\clearpage
\begin{figure}
\figurenum{\ref{fig:sfh}}
\plotone{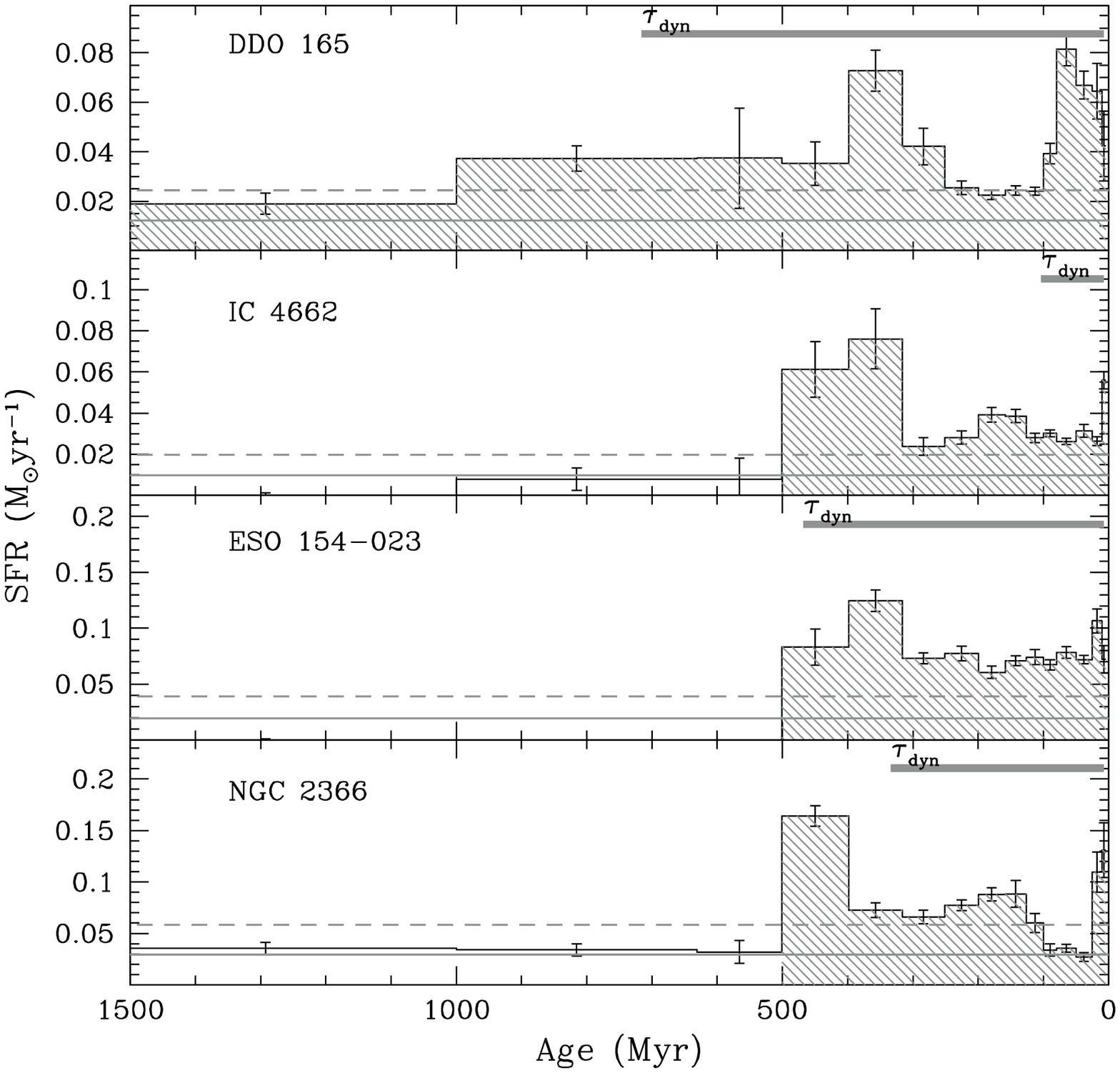}
\caption{\textit{Continued}}
\end{figure}

\clearpage
\begin{figure}
\figurenum{\ref{fig:sfh}}
\plotone{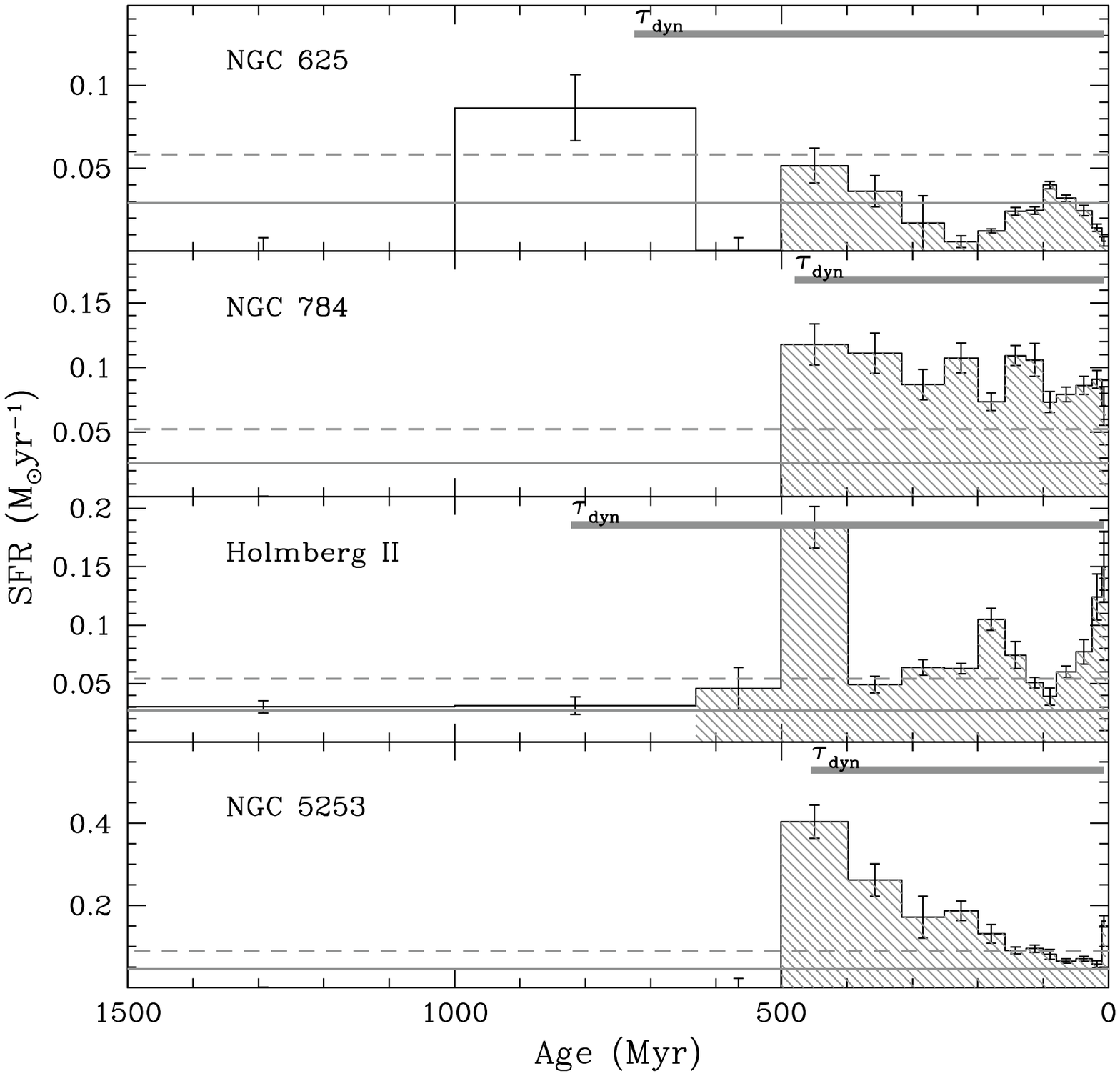}
\caption{\textit{Continued}}
\end{figure}

\clearpage
\begin{figure}
\figurenum{\ref{fig:sfh}}
\plotone{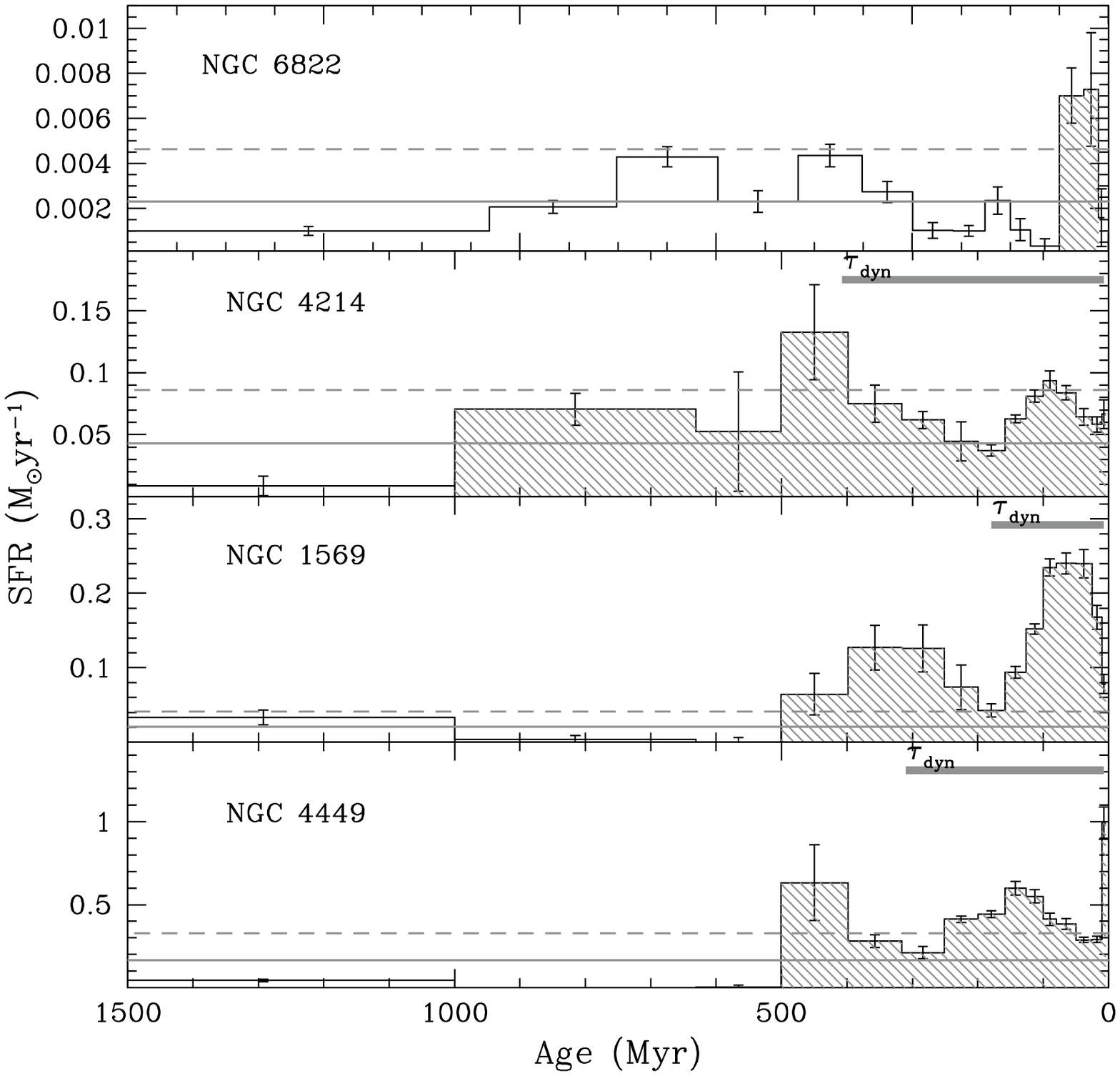}
\caption{\textit{Continued}}
\end{figure}

\clearpage
\begin{figure}
\plotone{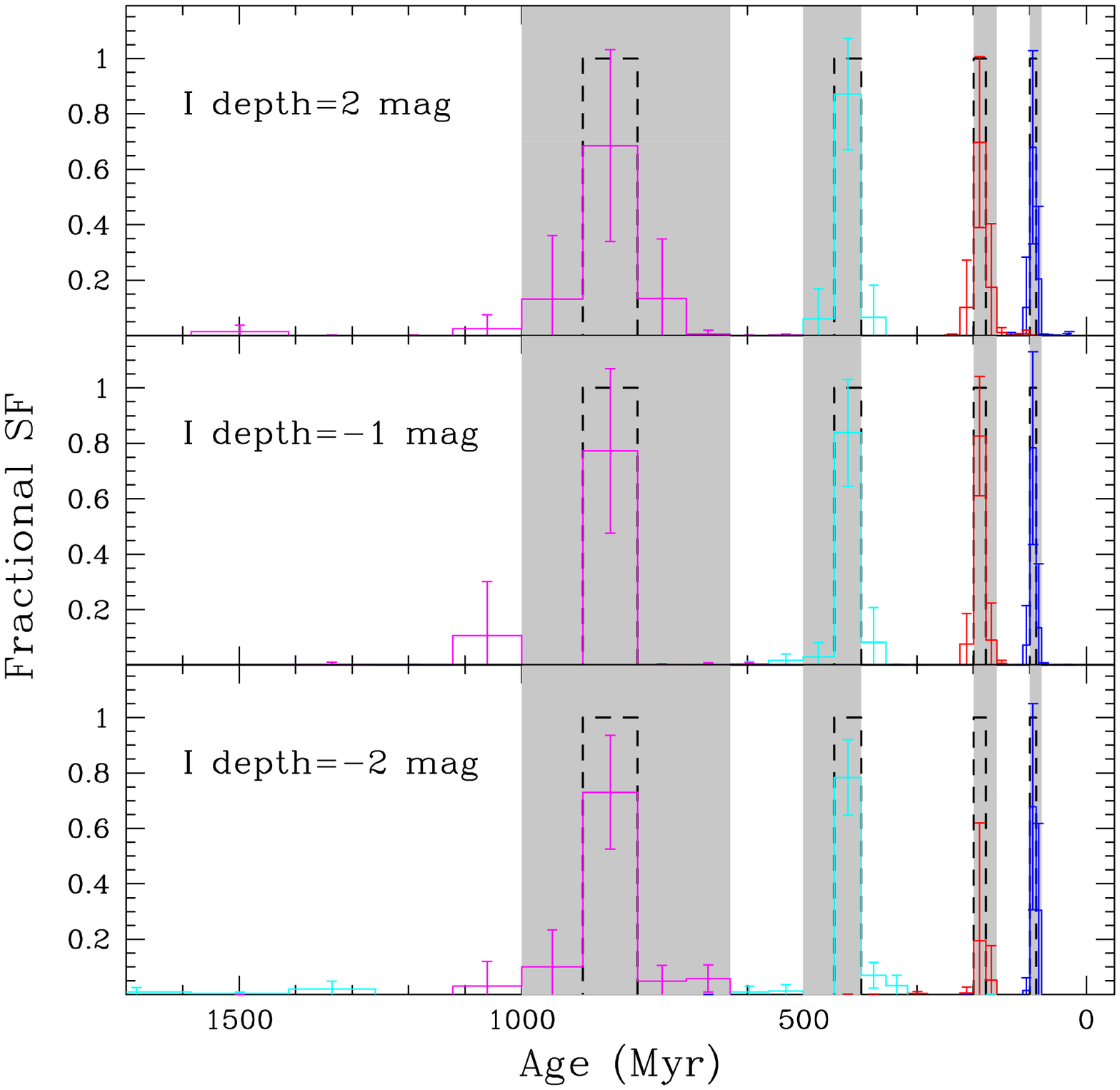}
\caption{Using SF input at varying look-back times (dashed lines at 100 Myr, 200 Myr, 450 Myr, 815 Myr), we recovered the SFHs from each single age populations (blue, red, cyan, and magenta lines) from photometry reaching a depth of I$\sim2$ mag (i.e., Antlia dwarf galaxy, top panel), I$\sim-1$ mag (i.e., NGC~4068, typical of most of our sample, middle panel), and I$\sim-2$ mag (i.e, NGC~4214, bottom panel). Our choice in time binning at the different look-back times is shown as shaded gray areas. The inherent time resolution is finer at recent times whereas the co-variance between the time bins at intermediate and ancient times justifies a coarser time resolution.}
\label{fig:timeres}
\end{figure}

\begin{figure}
\plotone{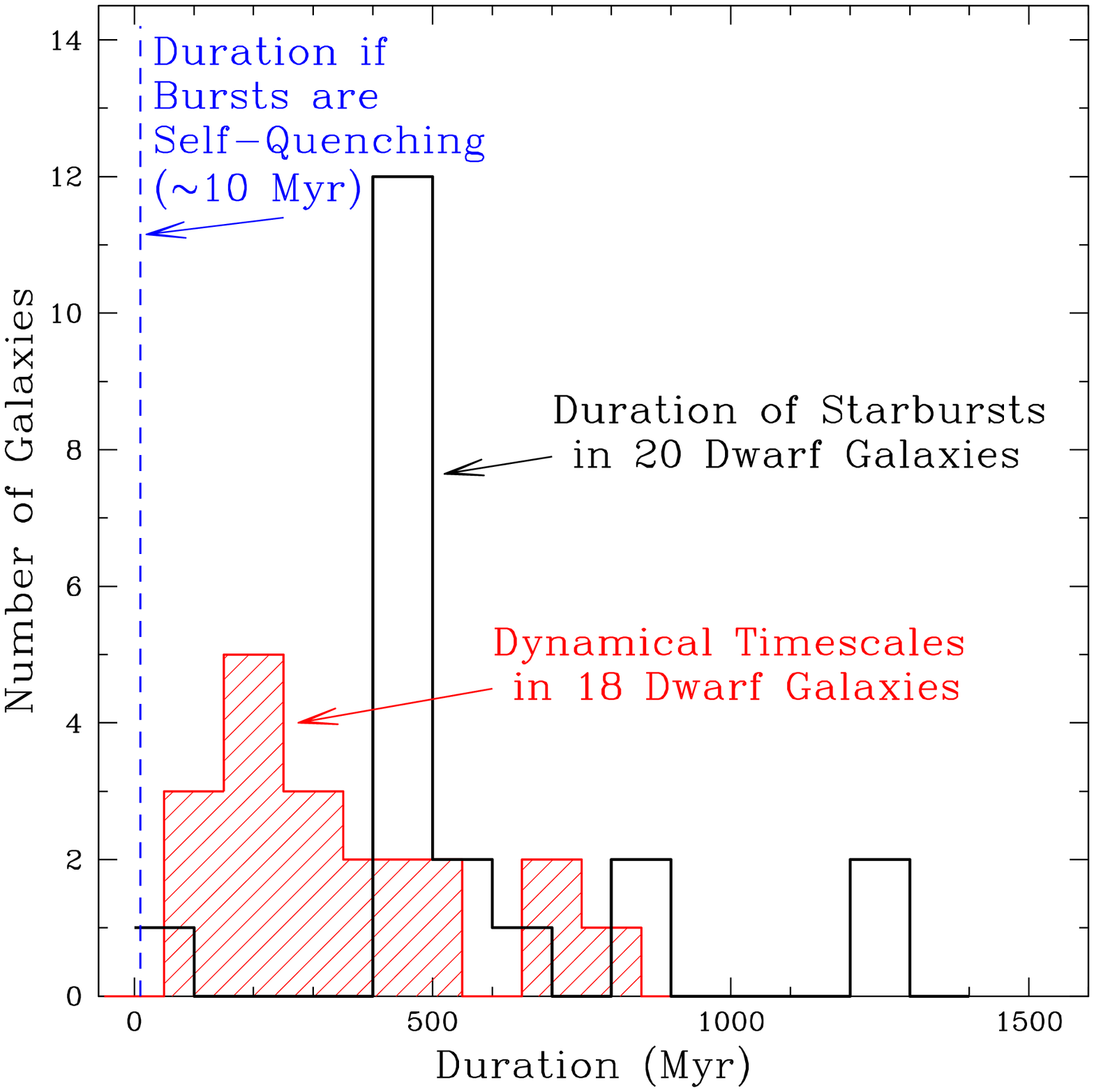}
\caption{The histogram shows the distribution of the starburst durations for the galaxy sample. The histogram bins are a simple scheme with each bin having the width of 100 Myr. The actual uncertainties in the measured durations vary; the shortest durations have the smallest uncertainty (i.e., $450\pm50$ Myr) and the longest durations have the largest uncertainty (i.e., $1300\pm290$ Myr). Also plotted is the distribution of the dynamical timescales for the seventeen galaxies for which rotation speeds are available. Interestingly, the duration of the starbursts for fifteen of the galaxies lie in the range $\sim450- \sim650$ Myr with the majority presenting burst duration of 450 Myr. Many of these durations are lower limits as the bursts are ongoing; the distribution would likely shift to even longer timescales with an exclusively fossil burst sample. Two galaxies in the sample show a longer duration greater than 1 Gyr. The timescale for the evolution of massive stars associated with ``self-quenching'' is plotted to show that the scale of these burst durations far exceed this timescale.}
\label{fig:histo_durations}
\end{figure}


\begin{thebibliography}{}
\bibitem[Calzetti et al.(1997)]{Calzetti1997} Calzetti, D., Meurer, 
G.~R., Bohlin, R.~C., Garnett, D.~R., Kinney, A.~L., Leitherer, C., 
\& Storchi-Bergmann, T.\ 1997, \aj, 114, 1834 
\bibitem[Cannon et al.(2003)]{Cannon2003} Cannon, J.~M., 
Dohm-Palmer, R.~C., Skillman, E.~D., Bomans, D.~J., C{\^o}t{\'e}, S., 
\& Miller, B.~W.\ 2003, \aj, 126, 2806 
\bibitem[Cole et al.(2007)]{Cole2007} Cole, A.~A., et al.\ 2007, 
\apjl, 659, L17 
\bibitem[Comins(1983)]{Comins1983} Comins, N.~F.\ 1983, \apj, 266, 
543 
\bibitem[Dalcanton(2007)]{Dalcanton2007} Dalcanton, J.~J.\ 2007, 
\apj, 658, 941 
\bibitem[Davies \& Phillipps (1988)]{Davies1988} Davies, J.~I. \& Phillipps, S., 1988, \mnras, 233, 553
\bibitem[Dekel \& Silk(1986)]{Dekel1986} Dekel, A., \& Silk, J.\ 1986, \apj, 303, 39
\bibitem[Dohm-Palmer et al.(2002)]{Dohm-Palmer2002} Dohm-Palmer, R.~C., 
Skillman, E.~D., Mateo, M., Saha, A., Dolphin, A., Tolstoy, E., Gallagher, 
J.~S., \& Cole, A.~A.\ 2002, \aj, 123, 813 
\bibitem[Dolphin(2000)]{Dolphin2000} Dolphin, A.~E.\ 2000, \pasp, 
112, 1383 
\bibitem[Dolphin(2002)]{Dolphin2002} Dolphin, A.~E., 2002, \mnras, 332, 91
\bibitem[Ferguson \& Babul(1998)]{Ferguson1998} Ferguson, H.~C. \& Babul, A., 1998, \mnras, 296, 585
\bibitem[Gallart et al.(2005)]{Gallart2005} Gallart, C., Zoccali, M., \& Aparicio, A.\ 2005, \araa, 43, 387 
\bibitem[Gerola et al.(1980)]{Gerola1980} Gerola, H., Seiden, 
P.~E., \& Schulman, L.~S.\ 1980, \apj, 242, 517 
\bibitem[Greggio et al.(1998)]{Greggio1998} Greggio, L., Tosi, M., Clampin, M., de Marchi, G., Leitherer, C., Nota, A. \& Sirianni, M., 1998, \apj, 504, 725
\bibitem[Harris et al.(2004)]{Harris2004} Harris, J., Calzetti, D., Gallagher, J.~S., Smith, D.~A. \& Conselice, C.~J., 2004, \apj, 603, 503
\bibitem[Heckman(2005)]{Heckman2005} Heckman, T., 2005, ASSL Volume 329, Starbursts from 30 Doradus to Lyman Break Galaxies, eds., De Gris, R. \& Gonz\'{a}lez Delgado, R.~M., Springer (the Netherlands), 3
\bibitem[Hogg \& Phinney(1997)]{Hogg1997} Hogg, D.~W. \& Phinney, E.~S., 1997, ApJL, 488, L95
\bibitem[Holmberg(1958)]{Holmberg1958} Holmberg, E.\ 1958, 
Meddelanden fran Lunds Astronomiska Observatorium Serie II, 136, 1 
\bibitem[Kauffmann et al.(1994)]{Kauffmann1994} Kauffmann, G., 
Guiderdoni, B., \& White, S.~D.~M.\ 1994, \mnras, 267, 981 
\bibitem[Kaviraj et al.(2007)]{Kaviraj2007} Kaviraj, S., Kirkby, 
L.~A., Silk, J., \& Sarzi, M.\ 2007, \mnras, 382, 960 
\bibitem[Kennicutt(1998)]{Kennicutt1998} Kennicutt, R.~C., Jr., 1998, \araa, 36, 189 
\bibitem[Kennicutt et al.(2005)]{Kennicutt2005} Kennicutt, R.~C., Jr., Lee, J.~C., Funes, J.~G., Sakai, S., \& Akiyama, S., 2005, ASSL Volume 329, Starbursts from 30 Doradus to Lyman Break Galaxies, eds., De Gris, R. \& Gonz\'{a}lez Delgado, R.~M., Springer (the Netherlands), 187
\bibitem[Krueger et al.(1995)]{Krueger1995} Krueger, H., Fritze-v.~Alvensleben, U., \& Loose, H.-H.\ 1995, \aap, 303, 41 
\bibitem[Larsen et al.(2008)]{Larsen2008} Larsen, S.~S., Origlia, 
L., Brodie, J., \& Gallagher, J.~S.\ 2008, \mnras, 383, 263 
\bibitem[Lee et al.(2009)]{Lee2009} Lee, J.~C., Kennicutt, 
R.~C., Jos{\'e} G.~Funes, S.~J., Sakai, S., \& Akiyama, S.\ 2009, \apj, 692, 1305 
\bibitem[Leitherer et al.(1999)]{Leitherer1999} Leitherer, C., et 
al.\ 1999, \apjs, 123, 3 
\bibitem[Loose \& Thuan(1986)]{Loose1986} Loose, H.-H., \& Thuan, T. X., 1986, Star-forming Dwarf Galaxies and Related Objects, ed. D. Kunth, T. X. Thuan, \& J. T. T. Van (Gif-sur-Yvette: Editions Frontires), 73
\bibitem[Mac Low \& Ferrara(1999)]{MacLow1999} Mac Low, M.-M., \& Ferrara, A.\ 1999, \apj, 513, 142 
\bibitem[Marigo \& Girardi(2007)]{Marigo2007} Marigo, P., \& Girardi, L.\ 2007, \aap, 469, 239 
\bibitem[Martin et al.(2002)]{Martin2002} Martin, C.~L., Kobulnicky, H.~A., \& Heckman, T.~M.\ 2002, \apj, 574, 663 
\bibitem[Mas-Hesse \& Kunth(1999)]{Mas-Hesse1999}Mas-Hesse, J. M. \& Kunth, D., 1999, A\&A, 349, 765
\bibitem[Mateo(1998)]{Mateo1998} Mateo, M.~L.\ 1998, \araa, 36, 435 
\bibitem[Mayer et al.(2001a)]{Mayer2001a} Mayer, L., Governato, F.,
Colpi, M., Moore, B., Quinn, T., Wadsley, J., Stadel, J., \& Lake, G.\
2001, \apjl, 547, L123
\bibitem[Mayer et al.(2001b)]{Mayer2001b} Mayer, L., Governato, F.,
Colpi, M., Moore, B., Quinn, T., Wadsley, J., Stadel, J., \& Lake, G.\
2001, \apj, 559, 754
\bibitem[Mayer et al.(2006)]{Mayer2006} Mayer, L., Mastropietro,
C., Wadsley, J., Stadel, J., \& Moore, B.\ 2006, \mnras, 369, 1021
\bibitem[McQuinn et al.(2009)]{McQuinn2009} McQuinn, K.~B.~W., 
Skillman, E.~D., Cannon, J.~M., Dalcanton, J.~J., Dolphin, A., Stark, D., 
\& Weisz, D.\ 2009, \apj, 695, 561 
\bibitem[McQuinn et al.(2010)]{McQuinn2010} McQuinn, K.~B.~W., Skillman, E.~D., Cannon, J.~M., Dalcanton, J.~J., Dolphin, A., Hidalgo-Rodriguez, S., Holtzman, J., Stark, D. et al., 2010, ApJ, in press
\bibitem[Meurer et al.(1997)]{Meurer1997} Meurer, G.~R., Heckman, 
T.~M., Lehnert, M.~D., Leitherer, C., \& Lowenthal, J.\ 1997, \aj, 114, 54 
\bibitem[Meurer(2000)]{Meurer2000} Meurer, G.~R., 2000, ASP Conf. Ser. Volume 211, Massive Stellar Clusters, ed. A. Lan\c{c}on, \& C. Boily (San Francisco: ASP), 81
\bibitem[Nishi \& Tashiro(2000)]{Nishi2000} Nishi, R., \& Tashiro, M.\ 2000, \apj, 537, 50 
\bibitem[Omukai \& Nishi(1999)]{Omukai1999} Omukai, K., \& Nishi, R.\ 1999, \apj, 518, 64 
\bibitem[Paturel et al.(2003)]{Paturel2003} Paturel, G., Theureau, G., Bottinelli, L., Gouguenheim, L., Coudreau-Durand, N., Hallet, N., \& Petit, C.\ 2003, \aap, 412, 57 
\bibitem[Romano et al.(2006)]{Romano2006}Romano, D., Tosi, M. \& Matteucci, F., 2006, \mnras, 365, 759
\bibitem[Scalo(1986)]{Scalo1986} Scalo, J.~M.\ 1986, Fundamentals 
of Cosmic Physics, Volume 11, 1
\bibitem[Schaerer, Contini, \& Kunth(1999)]{Schaerer1999} Schaerer, D., Contini, T., \& Kunth, D., 1999, A\&A, 341, 399
\bibitem[Schlegel et al.(1998)]{Schlegel1998} Schlegel, D.~J., 
Finkbeiner, D.~P., \& Davis, M.\ 1998, \apj, 500, 525 
\bibitem[Schulte-Ladbeck et al.(2001)]{Schulte-Ladbeck2001} 
Schulte-Ladbeck, R.~E., Hopp, U., Greggio, L., Crone, M.~M., 
\& Drozdovsky, I.~O.\ 2001, \aj, 121, 3007 
\bibitem[Searle et al.(1973)]{Searle1973} Searle, L., Sargent, 
W.~L.~W., \& Bagnuolo, W.~G.\ 1973, \apj, 179, 427
\bibitem[Silk et al.(1987)]{Silk1987} Silk, J., Wyse, R.~F.~G., \& Shields, G.~A., 1987, \apj, 322, L59
\bibitem[Skillman et al.(1988)]{Skillman1988} Skillman, E.~D., Terlevich, R., Teuben, P.~J., \& van Woerden, H.\ 1988, \aap, 198, 33 
\bibitem[Skillman et al.(1989)]{Skillman1989} Skillman, E.~D., 
Kennicutt, R.~C., \& Hodge, P.~W.\ 1989, \apj, 347, 875 
\bibitem[Skillman(1996)]{Skillman1996} Skillman, E.~D.\ 1996, The 
Minnesota Lectures on Extragalactic Neutral Hydrogen, 106, 208 
\bibitem[Skillman(1997)]{Skillman1997} Skillman, E.~D.\ 1997, 
Revista Mexicana de Astronomia y Astrofisica Conference Series, 6, 36 
\bibitem[Smith \& Gallagher(2000)]{Smith2000} Smith, L.~J., \& Gallagher, J.~S., III, 2000, ASP Conf. Ser. Volume 211,  Massive Stellar Clusters, ed. A. Lan\c{c}on, \& C. Boily (San Francisco: ASP), 90 
\bibitem[Spaans \& Norman(1997)]{Spaans1997} Spaans, M. \& Norman, C.~A., 1997, \apj, 483, 87
\bibitem[Stinson et al.(2007)]{Stinson2007} Stinson, G.~S., 
Dalcanton, J.~J., Quinn, T., Kaufmann, T., \& Wadsley, J.\ 2007, \apj, 667, 170 
\bibitem[Telles(2009)]{Telles2009} Telles, E.\ 2009, 
arXiv:0908.2966, To appear in ASP Series; Galaxy Wars
\bibitem[Thornley et al.(2000)]{Thornley2000} Thornley, M.~D., Schreiber, N.~M.~F., Lutz, D., Genzel, R., Spoon, H.~W.~W. \& Kunze, D., 2000, \apj, 539, 641
\bibitem[Tolstoy et al.(2009)]{Tolstoy2009} Tolstoy, E., Hill, V., \& Tosi, M.\ 2009, \araa, 47, 371 
\bibitem[Tosi et al.(1989)]{Tosi1989} Tosi, M., Greggio, L., \& Focardi, P.\ 1989, \apss, 156, 295 
\bibitem[Tremonti et al.(2001)] {Tremonti2001} Tremonti, C.~A., Calzetti, D., Leitherer, C. \& Heckman, T.~M., 2001, \apj, 555, 322 
\bibitem[Tully et al.(2006)]{Tully2006} Tully, R.~B., et al.\ 2006, \aj, 132, 729 
\bibitem[van Zee et al.(2004)]{vanZee2004} van Zee, L., Skillman,
E.~D., \& Haynes, M.~P.\ 2004, \aj, 128, 121
\bibitem[V{\'a}zquez \& Leitherer(2005)]{Vazquez2005} V{\'a}zquez, G.~A., \& Leitherer, C.\ 2005, \apj, 621, 695 
\bibitem[Weisz et al.(2008)]{Weisz2008} Weisz, D.~R., Skillman, E.~D., Cannon, J.~M., Dolphin, A.~E., Kennicutt, R.~C., Jr., Lee, J., 
\& Walter, F.\ 2008, \apj, 689, 160 
\bibitem[Weisz et al.(2009)]{Weisz2009} Weisz, D.~R., Skillman, 
E.~D., Cannon, J.~M., Dolphin, A.~E., Kennicutt, R.~C., Lee, J., 
\& Walter, F.\ 2009, \apj, 704, 1538 
\bibitem[Weldrake et al.(2003)]{Weldrake2003} Weldrake, D.~T.~F., de 
Blok, W.~J.~G., \& Walter, F.\ 2003, \mnras, 340, 12 
\end{thebibliography}
\end{document}